\documentclass[twocolumn, switch]{article}
\usepackage{preprint}
\usepackage{hyperref}
\usepackage[numbers,square]{natbib}

\RequirePackage[labelformat=simple]{subcaption}                                
\hypersetup{colorlinks=true, linkcolor=purple, urlcolor=blue, citecolor=cyan, anchorcolor=black}

\usepackage[utf8]{inputenc}	% allow utf-8 input
\usepackage[T1]{fontenc}
\usepackage{xcolor}
\usepackage{lineno}					% Line numbers
\usepackage{tikz} 					% ORCiD insertion

\usepackage{newfloat}
\DeclareFloatingEnvironment[name={Supplementary Figure}]{suppfigure}
\usepackage{sidecap}
\sidecaptionvpos{figure}{c}

\usepackage{titlesec}
\titlespacing\section{0pt}{12pt plus 3pt minus 3pt}{1pt plus 1pt minus 1pt}
\titlespacing\subsection{0pt}{10pt plus 3pt minus 3pt}{1pt plus 1pt minus 1pt}
\titlespacing\subsubsection{0pt}{8pt plus 3pt minus 3pt}{1pt plus 1pt minus 1pt}

\definecolor{lime}{HTML}{A6CE39}

\title{Burnup Measurement using Bent Crystal Diffraction Spectrometers for Pebble Bed Reactors}

\usepackage{amssymb}
\usepackage{amsmath}

\usepackage{authblk}

\author
	[1,*]{Ian Kolaja}

\author[1,2]{Lee Bernstein}

\author[1]{Ludovic Jantzen}

\author[1]{Eleanor Tubman}

\author[1]{Tatiana Siaraferas}

\author[1]{Massimiliano Fratoni}

\affil[1]{Nuclear Engineering Department, University of California Berkeley}
\affil[2]{Lawrence Berkeley National Laboratory}
\affil[*]{Corresponding author email address: ikolaja@berkeley.edu}

\begin{document}

\twocolumn[\begin{@twocolumnfalse}

\maketitle

\begin{abstract}
Burnup measurement is essential for monitoring and operating pebble-bed reactors (PBRs), where fuel pebbles circulate rapidly through the core. However, conventional gamma spectroscopy using high-purity germanium (HPGe) detectors is challenging due to high activity levels in discharge pebbles, leading to excessive dead time and Compton scattering. This study explores the use of bent crystal diffraction (BCD) spectrometers to filter the emitted gamma spectrum and isolate key peaks for improved measurement accuracy and speed. Pebble-wise depletion calculations were performed and the resulting spectra were analyzed using ray tracing (SHADOW3) and gamma response modeling (GADRAS). Key isotopes—$^{137m}$Ba/$^{137}$Cs, $^{239}$Np, $^{144}$Ce, $^{148m}$Pm, and $^{140}$La—were found to strongly correlate with burnup, residence time, core passes, plutonium production, and fluence. Machine learning regression models that were given synthetic spectra achieved a coefficient of determination ($R^2$) as high as 0.995 for burnup prediction. Among various BCD configurations, mosaic silicon crystals in the (440) orientation combined with an HPGe detector provided optimal performance for measuring $^{137}$Cs decay (via $^{137m}$Ba), while silicon (220) and (440) paired with scintillators were effective for the shorter-lived isotopes.
\end{abstract}


\vspace{0.5cm}

\end{@twocolumnfalse}]

\section{Introduction}

Pebble bed reactors (PBRs) operate on a continuous refueling mode that relies on a burnup measurement system (BUMs) capable of inferring the burnup level of each pebble~\cite{analysis-of-peb-bu-in-PBR}. The BUMs must process pebbles quickly to reduce the fuel inventory necessary to operate the reactor at any time and to be accurate to optimize fuel utilization and reduce operational uncertainties. An optimized BUMs could measure pebbles as fast as they leave the core, which is about 22.5 seconds for the the generic fluoride-salt-cooled high-temperature reactor (gFHR) design published by Kairos Power~\cite{gfhr}. Currently, the expectation is to use high-purity germanium (HPGe) detectors for assessing the concentration of gamma-emitting fission products, such as $^{137}$Cs, which are used as proxy for burnup~\cite{best-candidate-isotopes-burnup}. HPGe detectors are preferred for their excellent energy resolution. However, pebbles leaving the core are highly radioactive and the gamma spectrum for discharge pebbles show total count rates on the order of tens to hundreds of gigacounts per second (Gcps)~\cite{sandia-pbr-mca-kovacic-report}. Such extreme activity makes it difficult to operate HPGe detectors without increased cooling time and substantial shielding or detector distance. In conventional configurations, HPGe detectors are typically limited to count rates of only a few tens of thousands per second, and specialized designs are required to approach even 1 Gcps~\cite{highcounthpge}. Additionally, Compton scattering and nearby peaks can obscure the 661 keV photopeak from $^{137}$Cs~\cite{areva-readiness-study}. 

This work proposes to use bent crystal diffraction (BCD) spectrometer for burnup measurements, a technique that is widely applied in nuclear physics, astrophysics, synchrotron science, and nuclear forensics~\cite{germanium-bent-monochromator-tungsten-levels-dumond, bcd-solar-flare-nasa-sylwester, synchrotron-two-bcd-monochromators-yamaoka, bcd-plutonium-spectrometer-goodsell}. BCD spectrometers can filter gammas by energy before they reach the detector. In this method, precisely fabricated crystals act as diffraction gratings for collimated gamma radiation. The diffraction angle depends on the photon wavelength, which is directly related to the gamma energy as shown in Equation~\ref{eqn:waveenergy}.
\begin{equation}
  E\lambda=hc
  \label{eqn:waveenergy}
\end{equation}
When Bragg's law is satisfied with respect to the incident angle and energy of the photons, constructive diffraction occurs as given by Equation~\ref{eqn:braggslaw}. 
\begin{equation}
  E=\frac{hc}{2}\frac{n}{d}\frac{1}{sin\theta}
  \label{eqn:braggslaw}
\end{equation}
The crystal is bent to improve efficiency by maintaining the correct Bragg angle across the crystal surface. Perfect monocrystals are typically used to maximize energy resolution. Mosaic crystals, composed of slightly misaligned crystallites, enhance efficiency by relaxing the Bragg condition \cite{xray-diffraction-theory-book-zachariasen}, increasing the diffracted flux by up to two orders of magnitude~\cite{mosaic-crystal-monochromators-freund}.

In a BUMS using a BCD spectrometer, the detector is positioned along the diffracted beam path at an angle of $2\theta$. The BCD spectrometer thus functions as an energy band-pass filter for gamma rays incident on the detector. The bandwidth can be further reduced by placing shielding with a narrow slit in front of the detector. In certain configurations, this enables energy resolution superior to that of an HPGe detector. If the spectrometer isolates a single gamma, the detector’s intrinsic resolution becomes irrelevant, allowing lower-cost detectors—such as scintillators—to be used instead.

This approach has practical limitations. Adjusting the filtered energy range requires precise repositioning of the crystal or source, which is impractical under the rapid measurement conditions of PBR operation. Furthermore, the degree of collimation required for diffraction alignment significantly reduces photon throughput, necessitating a high source activity. Consequently, a static configuration that targets only the most intense and diagnostically useful gammas is preferred.

In this study, Serpent 2 is used to simulate the gFHR design and obtain pebble compositions and gamma spectra~\cite{serpent}. Candidate gamma rays for measurement are identified using data science techniques. Synthetic spectrometer measurements are simulated using SHADOW3~\cite{shadow3} and machine learning regression models are trained on this data and evaluated. GADRAS is used to visualize a filtered spectra observed with the detector~\cite{GADRAS}. This study is based on Chapter~2 of the author’s PhD dissertation~\cite{ian-dissertation}, completed under the supervision of Massimiliano Fratoni. Key data and analysis code used in this work have been archived on Zenodo~\cite{hxf-count-data, zenodo-code-archive}.

% MAX CHECK: Is this a good way of citing dissertation? 

\section{Synthetic Spectrum Generation}
\label{sec:spectrumgeneration}

\subsection{Pebble Data Generation with HxF}
\label{subsec:PebbleDataGeneration}

The HxF tool~\cite{hxf} is used to simulate the full irradiation history of individual pebbles by integrating the Monte Carlo neutron transport code Serpent 2.2.0~\cite{serpent} with a Python-based framework. HxF sequentially handles Serpent calculations (including transport and burnup), pebble motion, recirculation, burnup evaluation, and the removal and replacement of pebbles based on a user-defined discard threshold~\cite{hxf}. For each pebble, HxF outputs nuclide inventories and detailed history metrics.

HxF was used to generate pebble data for the gFHR model~\cite{gfhr} at equilibrium, operating at 100\% power with a 180 MWd/kgHM discard threshold. Unlike in the benchmark, pebbles positions are not determined by a discrete element method (DEM) simulation by rather arranged in a face-centered cubic (FCC) lattice. The FCC lattice parameter is set to $a = 2.98$~cm, achieving a packing fraction of 59.88\% with 250,190 pebbles. This structured placement enables a simplified, layer-by-layer vertical motion~\cite{hxf}. 

Pebbles with a higher burnup than the set threshold are not reinserted in the core. Under these equilibrium conditions, pebbles pass on average between 7 and 8 times through the core but make no more than 10 passes. Beyond burnup, the pebble history metrics tracked include residence time, average radial distance, the number of passes through the core, and fluence (both thermal and fast). These properties are tracked both cumulatively and from the pebble's most recent pass through the core. The dataset created for this study includes 15,744 discharged pebbles taken from 8 different time steps at equilibrium. The statistical error on the multiplication factor was below 50 pcm for these simulations. The pebble-wise values for power and flux, both thermal and fast, all have a statistical error under $\pm$5\%.

\begin{figure}[!htbp]
  \centering
  \includegraphics[width=\columnwidth]{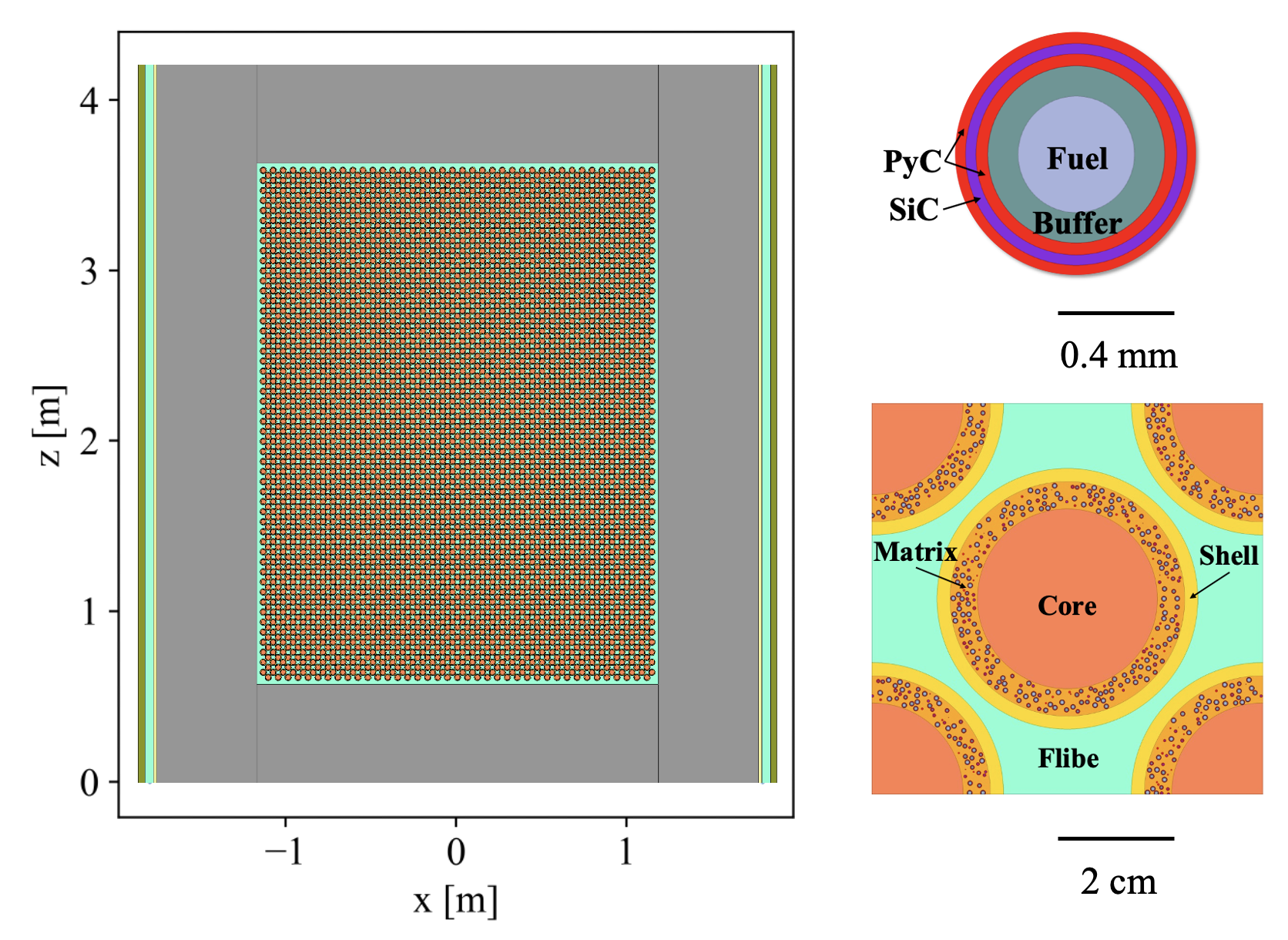}
  \caption{Serpent generated plot of the gFHR model in HxF, including the core geometry (left), explicit pebble (top right), and TRISO geometry (bottom right)~\cite{gfhr}.}
  \label{fig:hxf_demo}
\end{figure}

\subsection{Gamma Transmission to Pebble Surface}
\label{subsec:PebbleTransmission}
\begin{figure}[!htbp]
    \centering
    \includegraphics[width=0.9\columnwidth]{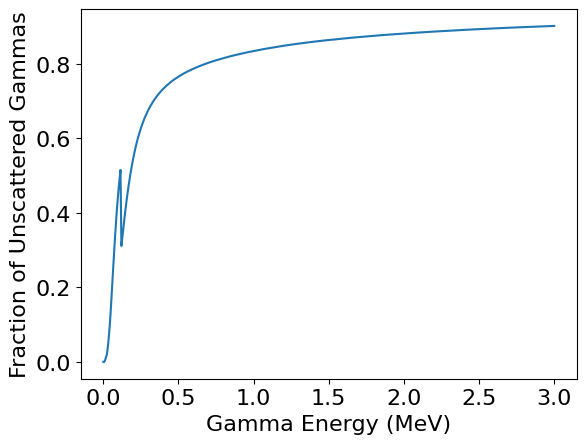}
    \caption{Fraction of gammas emitted from TRISO fuel kernels that exit the pebble's surface at full energy without being absorbed.}
    \label{fig:transmission_curve}
\end{figure}

Gammas, especially at lower energies, have a chance of being scattered or absorbed before they reach the outer surface of the pebble. Serpent was used to determine the fraction of gammas that reach the pebble surface without being absorbed or losing energy. Simulations were performed assuming a fresh fuel pebble in which a mono-energetic gamma source was sampled from the fuel kernels at 500 energies between 1.125 keV and 3000 keV. Each simulation had a tally tracking the outward current of full energy gammas, which is shown in Figure~\ref{fig:transmission_curve}. This energy-dependent factor will be called the escape efficiency, $\varepsilon _{escape}$, and can be interpolated at different energies to estimate the self-shielding effects of the fuel.

\subsection{Nuclide Selection}
\label{subsec:NuclideSelection}

Identifying nuclides that can realistically be measured while providing useful information is important for limiting the design space and cost. Since each spectrometer can only target a narrow energy range, additional nuclides will generally require an additional detector and crystal pair. Thus, gammas that satisfy the following criteria are considered:

\begin{enumerate}
\item They have a high enough source rate to allow sufficient counts even with low efficiency and short measurement times (over $10^{10}$ emissions per second).
\item They are emitted at between 30 keV and 1 MeV. Low energy gammas are penalized by interacting more with the pebble as they leave. However, higher energy gammas suffer from increasingly low diffraction efficiency and near zero Bragg Angles.
\item They are emitted from a nuclide whose concentration correlates with pebble history. Shorter lived nuclides (under 60 days), for example, give more information about recent power history.
\item Their energy allows them to be measured without significant interference from other nuclides in the form of either peak crowding or potentially higher order diffraction. 
\end{enumerate}

40 nuclides with decay gammas that satisfy criteria 1 and 2 were found. Data analysis techniques can provide additional insight on correlations between these nuclides and fuel history parameters. One approach to this involves using a random forest regression (RFR) model trained on the HxF nuclide concentrations as input (assumed to be perfectly known) and the target history parameters as output. RFR models, as implemented with scikit-learn~\cite{scikit}, have built in methods for calculating the feature importance of input features in units of Mean Decrease in Impurity (MDI). MDI quantifies the probability of reaching a certain node that is averaged over all trees in the forest model. RFRs are also well suited for tabular data sets with around 10,000 samples, and offer shorter training times, better performance before hyperparameters are trained, and more interpretability compared to deep learning models~\cite{tree-models-outperform-deep-learning-tabular-grinsztajn}. In addition to feature importance, these models also provide an upper limit of accuracy if the nuclides are measured perfectly.

\begin{figure*}[hp]
  \centering
  \begin{subfigure}[b]{0.35\textwidth}
    \centering
    \includegraphics[width=\textwidth]{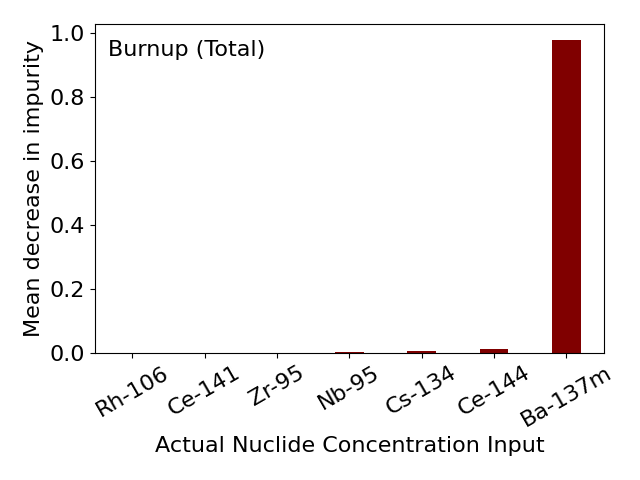}
  \end{subfigure}
  \begin{subfigure}[b]{0.35\textwidth}
    \centering
    \includegraphics[width=\textwidth]{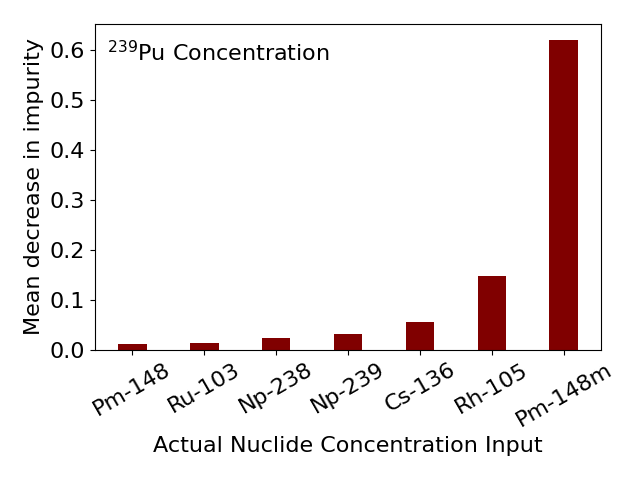}
  \end{subfigure}
  
  \begin{subfigure}[b]{0.35\textwidth}
    \centering
    \includegraphics[width=\textwidth]{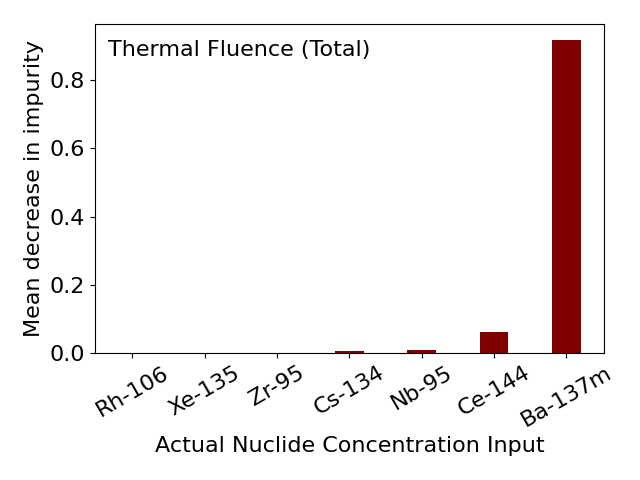}
  \end{subfigure}
  \begin{subfigure}[b]{0.35\textwidth}
    \centering
    \includegraphics[width=\textwidth]{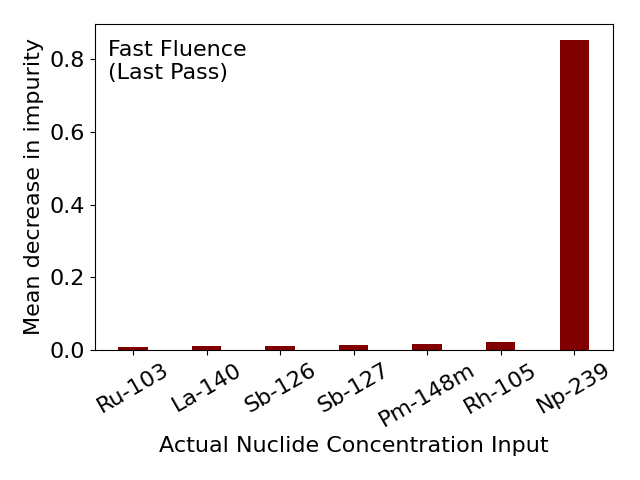}
  \end{subfigure}
  
  \begin{subfigure}[b]{0.35\textwidth}
    \centering
    \includegraphics[width=\textwidth]{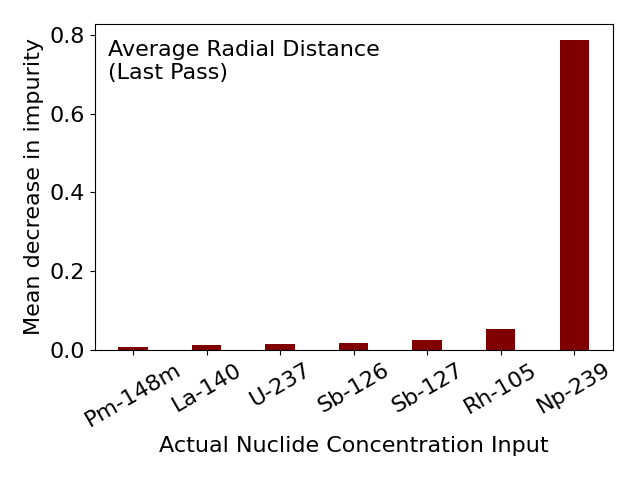}
  \end{subfigure}
  \begin{subfigure}[b]{0.35\textwidth}
    \centering
    \includegraphics[width=\textwidth]{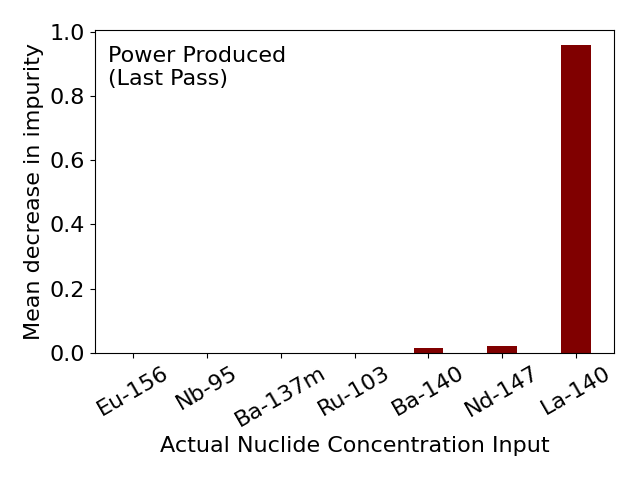}
  \end{subfigure}
  \caption{Calculated feature importance for six target variables: Total pebble burnup, the total concentration of $^{239}$Pu, the total thermal fluence experienced by the pebble, the fast fluence experienced by the pebble on its most recent pass, the average radial pathway of the pebble on its most recent pass, and the power produced by the pebble on its last pass.}
  \label{fig:featureimportance}
\end{figure*}

\begin{figure*}[ptb]
  \centering
  \begin{subfigure}[b]{0.35\textwidth}
    \centering
    \includegraphics[width=\textwidth]{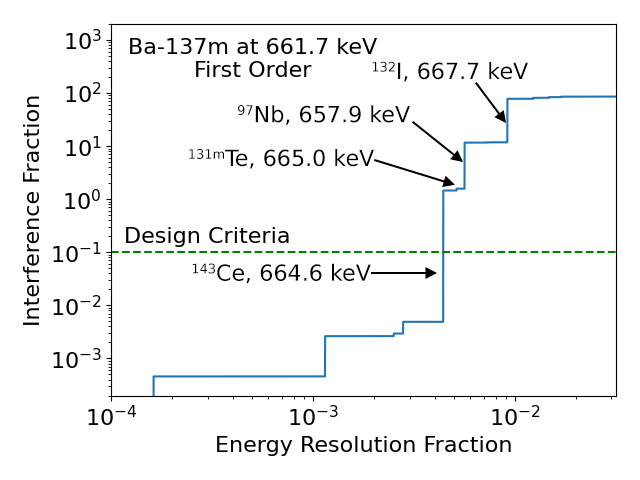}
  \end{subfigure}
  \begin{subfigure}[b]{0.35\textwidth}
    \centering
    \includegraphics[width=\textwidth]{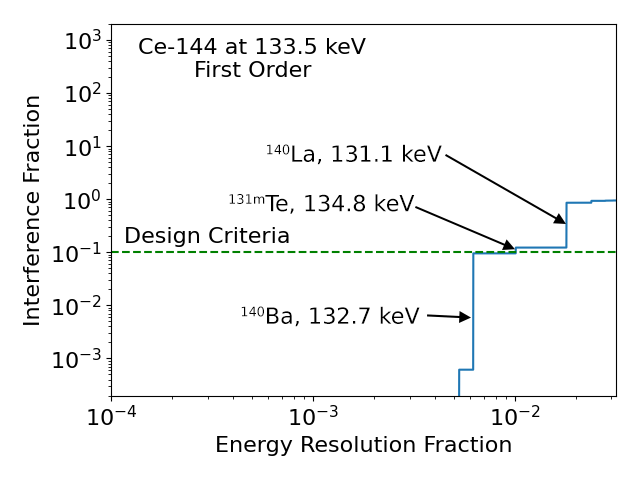}
  \end{subfigure}
  
  \begin{subfigure}[b]{0.35\textwidth}
    \centering
    \includegraphics[width=\textwidth]{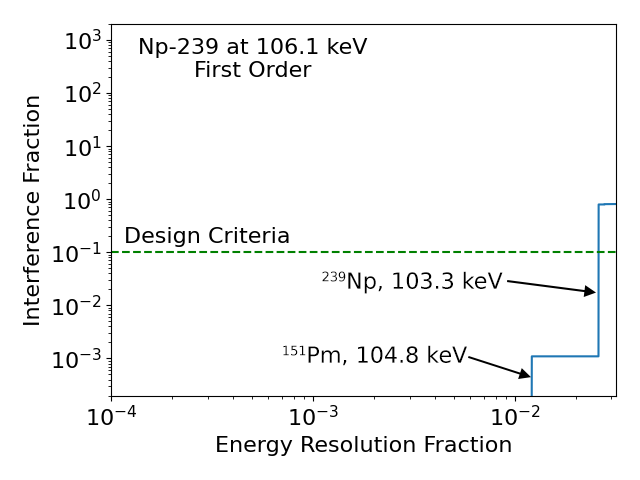}
  \end{subfigure}
  \begin{subfigure}[b]{0.35\textwidth}
    \centering
    \includegraphics[width=\textwidth]{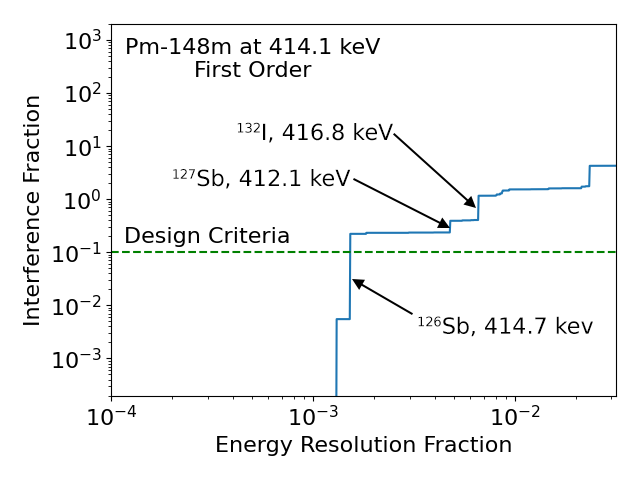}
  \end{subfigure}
  \caption{Plots showing the relationship between fraction of interference and the width of energies making it to the detector via diffraction. The points at which notable contributors to interference drop off are annotated. Targeted measured gammas include 661.7 keV from $^{137m}$Ba, 133.5 keV from $^{144}$Ce, 106.1 keV from $^{239}$Np, and 414.1 keV from $^{148m}$Pm.}
  \label{fig:interference_curves}
\end{figure*}

For several history parameters, the nuclides with the highest MDI are shown in Figure~\ref{fig:featureimportance}. The gamma spectra of the top nuclides were then investigated for specific gammas that meet all four criteria. The interference around the target peak was quantified using Equation~\ref{eqn:interference_fraction}, where $S_{targeted}$ is the source rate of the desired gamma and $S_{unwanted}$ is the emission rate of every other gamma within in the energy range $\Delta E$. 
\begin{equation}
  f_{interference}=\frac{\sum S_{unwanted}(\Delta E)}{S_{targeted}}
  \label{eqn:interference_fraction}
\end{equation}
This assumes the BCD spectrometer filters all gammas in the energy range uniformly, which is conservative. For a low energy resolution detector potentially being operated in current mode, the interference is thus a constant source of noise. For a detector with good energy resolution, "interfering" gammas could still be resolvable if their energy is different enough from the target gamma.

A starting target limit of 10\% interference is used. Interference was calculated for two cases: with just first-order diffraction considered, and with up to fourth order diffraction considered. Higher order diffraction would allow gammas whose energy is a multiple of the first order energy to reach the detector. The efficiency of higher order diffraction is often much lower, but here it is treated the same as a conservative estimate. It should be noted that most scintillators could resolve gammas from different diffraction orders. A 100 keV gamma that undergoes first order diffraction could likely be separated from a 200 keV undergoing second order diffraction. The main purpose of looking at the higher orders is largely to ensure that there are no higher intensity gammas that could mask the first order signal with Compton scattering or cause dead time. For a few nuclides being considered, the trade off between energy filter width and interference is shown in Figure~\ref{fig:interference_curves}. These plots show how narrow the filter must be to eliminate most interference in the worst case scenario. If the interference is higher than the design criteria for a wide range of energy resolution values, then the associated gamma likely cannot meet the fourth criteria. In this study, energy resolution is used interchangeably with the width of the spectrometer's energy filter.

\begin{table*}[htbp]
    \caption{Nuclear data for selected nuclei and gammas ~\cite{nuclear-data-a-137,nuclear-data-a-144,nuclear-data-a-239,nuclear-data-a-148,nuclear-data-a-140,nudat3}. Average emission rates are calculated from all fuel pebble samples.}
    \label{tab:nuclide_properties}
    \centering
    \begin{tabular}{|l|l|l|l|}\hline
        Parent Isotope & Selected Energy (keV) & Half Life & Typical Emission Rate (Bq) \\ \hline
        $^{137m}$Ba/$^{137}$Cs & 661.7  & 30.08y & $4\times10^{10}$ \\ \hline
        $^{144}$Ce & 133.5 & 284.9d & $2\times10^{10}$ \\ \hline
        $^{239}$Np & 106.1 & 2.36d & $2\times10^{12}$ \\ \hline
        $^{148m}$Pm & 414.07 & 41.29d & $1\times10^{10}$ \\ \hline
        $^{140}$La & 1,596 & 40.284h &  $3\times10^{12}$ \\ \hline
    \end{tabular}
\end{table*}

This process narrowed down the options for target nuclides and gammas. $^{137}$Cs is long-lived ($t_{1/2}=$30.08 y) fission product that is commonly used for burnup measurement. The decay of $^{137}$Cs, or more precisely the isomeric state of $^{137m}$Ba, is measured using its 661.7 keV gamma. With short fuel decay times, this gamma can be subject to significant crowding even before Compton scattering is considered. Potentially crowding gammas include the 667.7 keV gamma from $^{132}$I ($t_{1/2}=$2.28 h), the 657.9 keV gamma from $^{97}Nb$ ($t_{1/2}=$72.1 m), and the 664.6 keV gamma from $^{143}$Ce ($t_{1/2}=$33.037 h). An energy filter width of about 0.4\%, or 2.65 keV, is required to completely eliminate them. However, because 661 keV is relatively high energy, it is subject to lower diffraction efficiency; this makes achieving such an energy resolution likely impossible without sacrificing too much throughput. Thus, using a detector in current mode is not viable. However, using a higher efficiency spectrometer with a coarser filter in combination with a HPGe detector is viable. The $^{137m}$Ba peak can still be clearly resolved in this way without being masked by Compton scattering, and the nearby peaks could provide extra information for regression models.

$^{239}$Np ($t_{1/2}=$2.36d) is another prime candidate, since its decay produces $^{239}$Pu and it is correlated with short-term fission history. $^{239}$Np has many gamma lines that are intense, though their lower energy would make measuring them with conventional gamma spectroscopy nearly impossible due to the Compton backscattering peak. The 61.5 keV and 106.1 keV gammas both have high intensities with low interference. The emission rate of 106.1 keV is an order of magnitude higher, but a spectrometer designed for it will suffer from lower reflectivity and a lower Bragg angle. An energy resolution of about 2\% eliminates most interference for both gammas. However, the largest source of interference for the 106.1 keV gamma is actually the 103 keV X-ray that also originates from $^{239}$Np. Thus, the 106.1 keV gamma was selected.

$^{144}$Ce is the second most informative nuclide for burnup, likely due to its medium half life ($t_{1/2}=$284.9d). It has a 133.5 keV gamma with little interference at an energy resolution of 0.6\%. $^{148m}$Pm is informative for predicting the $^{239}$Pu concentration. This could be because it has a cumulative fission yield of $4.5\times10^{-8}$ in $^{239}$Pu and a markedly lower yield of ${4.8\times10^{-11}}$ in $^{235}U$~\cite{endf}. For its 414.1 keV gamma, an energy resolution of under 0.1\% is required to eliminate most interference. While $^{148m}$Pm has lower energy gammas, many of them face significant crowding or have lower intensity. Finally, $^{140}$La is a key nuclide for determining recent power history. Because its decay produces one of the highest energy and intensity gammas in the spectrum, it could easily be measured directly with a scintillator. Thus, its synthetic signal is included in the regression analysis, but a BCD spectrometer is not designed for it. Key nuclear properties for the targeted gammas and their parent nuclides are provided in Table~\ref{tab:nuclide_properties}.

\subsection{Bent Crystal Spectrometer Optical Modeling}
\label{subsec:DiffractionModel}
\begin{figure}[htb]
  \centering
  \includegraphics[width=0.9\columnwidth]{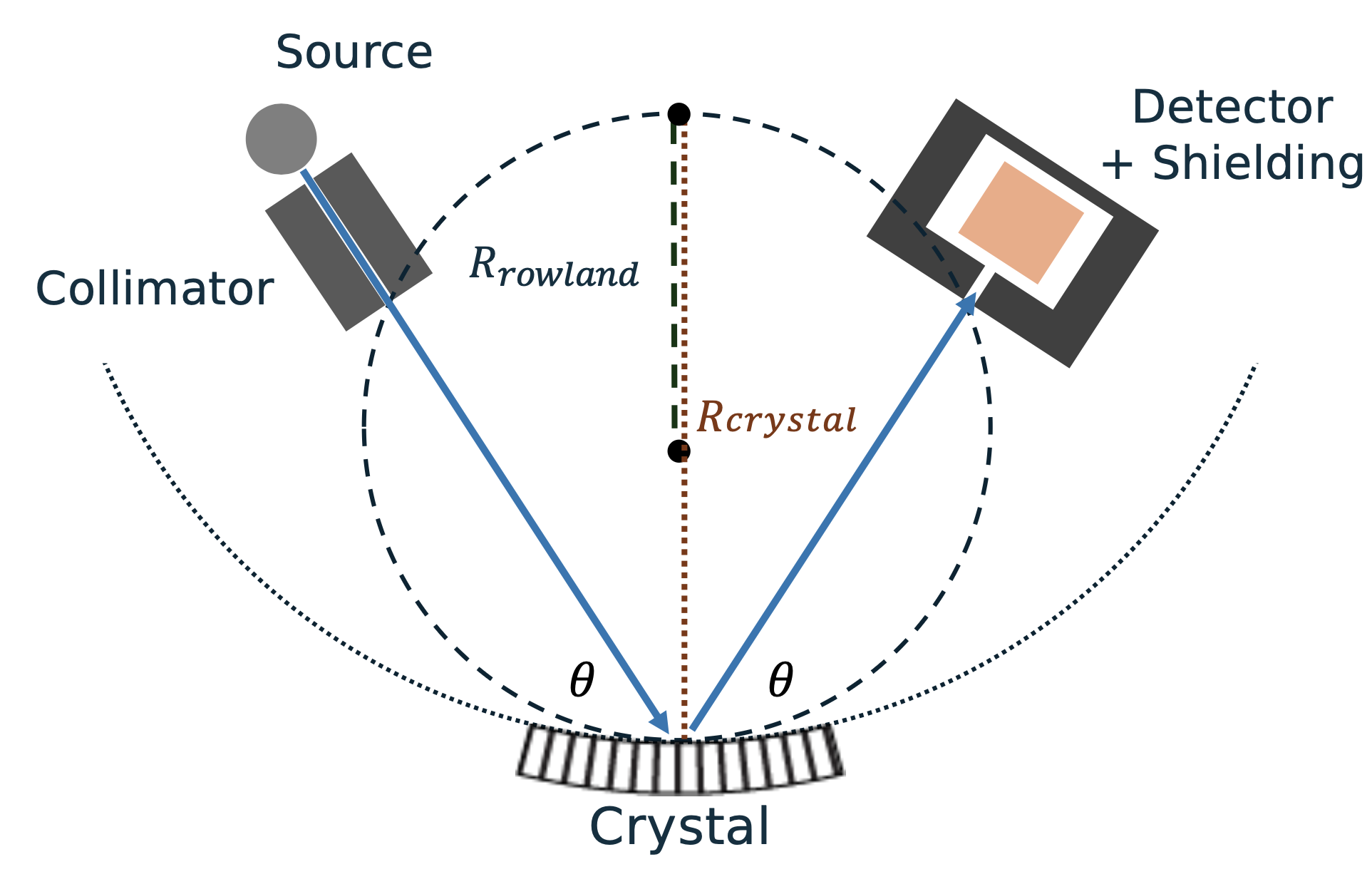}
  \caption{Illustration of BCD spectrometer in Bragg geometry. The source is a collimated fuel pebble, and the detector has a slit open towards the crystal.}
  \label{fig:bragg_diagram}
\end{figure}

SHADOW3 was used to model the optical performance of candidate BCD spectrometer designs using ray-tracing calculations~\cite{shadow3}. Though it is primarily used for the simulation of synchrotron beamlines, it has well developed crystal diffraction models. Symmetric Bragg geometry is used, placing both the source and detector at the same distance from the crystal and at an angle $\theta$ relative to the crystal's centerline~\cite{gamma-spectroscopy-first-85-years-deslattes}. The Rowland circle connects all three elements with radius $R_{rowland}$. The crystal is cylindrically bent with a radius of curvature, $R_{crystal}$, that is twice that of the Rowland circle. This setup eliminates aberrations while ensuring gammas that satisfy Bragg's law are focused on the detector. A circular source was used to represent a fuel pebble whose decay gammas travel through a borehole collimator with aperture $d$ and focal length $f$. The source was given a uniform half-width divergence, $\Delta_{src}/2$, which was estimated using Equation~\ref{eqn:collimatordivergence}. The setup is illustrated in Figure~\ref{fig:bragg_diagram}.
\begin{equation}
  \Delta_{src} \approx \frac{d}{f}
  \label{eqn:collimatordivergence}
\end{equation}

\begin{figure}[tb]
  \centering
  \begin{subfigure}[b]{0.9\columnwidth}
    \centering
    \includegraphics[width=\textwidth]{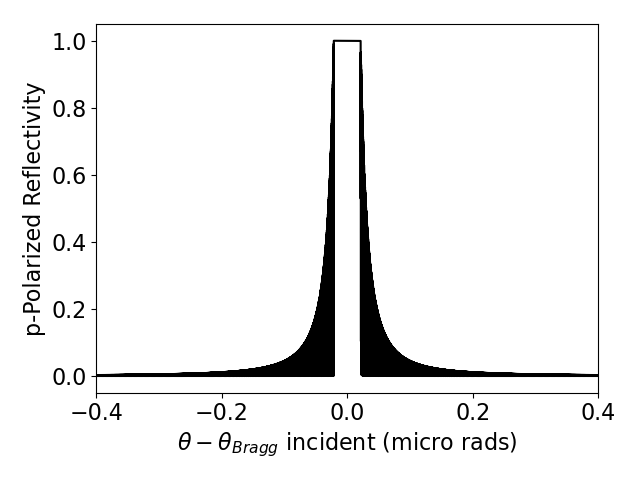}
  \end{subfigure}
  \begin{subfigure}[b]{0.9\columnwidth}
    \centering
    \includegraphics[width=\textwidth]{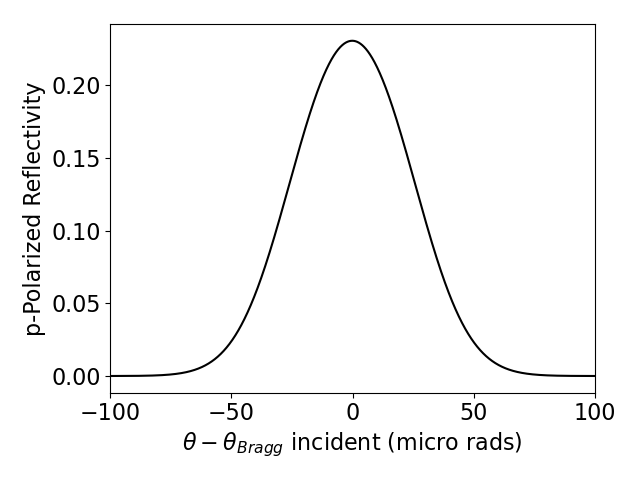}
  \end{subfigure}
  \caption{P-polarized reflectivity curve calculated by XOP for silicon (660) at 661 keV for a perfect crystal (top) and a 0.003° FWHM mosaic crystal (bottom). The greater width of the reflectivity curve for the mosaic crystal illustrates its significantly higher throughput at the cost of energy resolution.}
  \label{fig:reflectivity_curves}
\end{figure}

Each model was designed for the targeted gamma to undergo first order diffraction. The XOP (X-ray OPtics utilities) interface in SHADOW3 was used to determine the crystal reflectivity as a function of the incoming Bragg angle~\cite{xop}. This is demonstrated in Figure~\ref{fig:reflectivity_curves} for a perfect crystal and mosaic crystal. Reflectivity is strongly influenced by the gamma energy and the crystal plane and mosaicity.

\begin{figure}[!hbt]
    \centering
    \includegraphics[width=0.9\columnwidth]{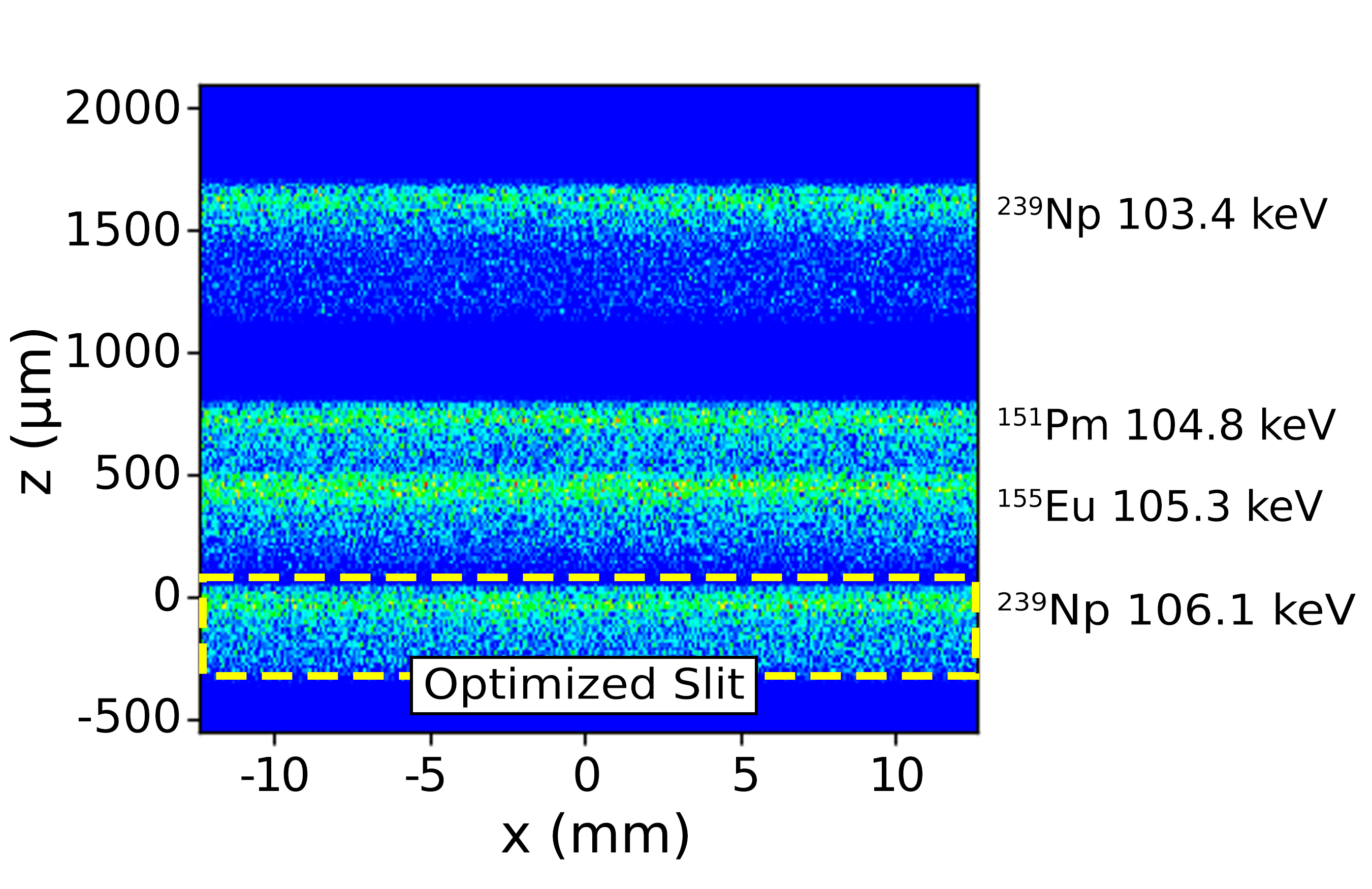}
    \caption{Example diffraction pattern incident on a detector for a BCD spectrometer setup targeting $^{239}$Np. For illustration purposes, lower mosaicity is used, and only gammas at four energies are simulated with equal intensity. The box around the $^{239}$Np line represents the unshielded part of the detector.}
    \label{fig:diffraction_patterns}
\end{figure}

In SHADOW3, the detector was simulated as a circular screen on the image plane. The diffraction efficiency was calculated by simulating 10,000,000 monoenergetic rays of the desired energy. The incident diffraction pattern on the detector was also produced; this is useful because shielding with an opening that matches the diffraction pattern of the targeted gamma can be placed in front of the detector, further blocking unwanted gammas. The energy resolution was determined by simulating four times as many rays with a source that has a uniform energy distribution centered on the target gamma energy. This allowed the Full Width Half Maximum (FWHM) to be determined in terms of both the energy band that diffracts in the crystal and the energy band accepted by the detector slit. An example of the diffraction pattern incident on a detector is shown in Figure~\ref{fig:diffraction_patterns}. The way the detector slit also narrows the energy filter by leveraging the angular dispersion of the gammas is also illustrated.

\subsection{Detector Geometry Optimization}
\label{subsec:geometry_optimization}

The crystal properties have the largest impact on the overall cost and performance of the spectrometer. Increasing the mosaicity of the crystal widens the reflectivity curve, increasing throughput while allowing more gamma energies to diffract. Using a crystal plane with smaller interplanar spacing increases the Bragg angle, which is important for higher energy gammas to avoid having a nearly straight beam. However, these crystal planes often have a lower reflectivity, and it can also be more expensive to acquire crystals manufactured in such orientations. Thus, it is preferable to use silicon (220) or (111) crystal planes when possible. For $^{239}$Np, silicon (220) is suitable, while higher-energy gammas from the other nuclides are better measured using silicon (440). Finally, increasing the diffracting crystal dimensions, particularly the thickness, increases both efficiency and cost. This can be achieved practically by stacking many smaller crystals.

\begin{figure*}[!bht]
  \centering
  \begin{subfigure}[b]{0.41\textwidth}
    \centering
    \includegraphics[width=\textwidth]{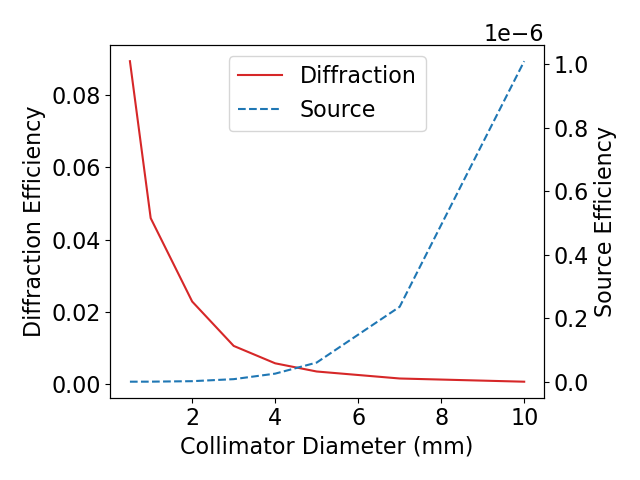}
  \end{subfigure}
  \begin{subfigure}[b]{0.41\textwidth}
    \centering
    \includegraphics[width=\textwidth]{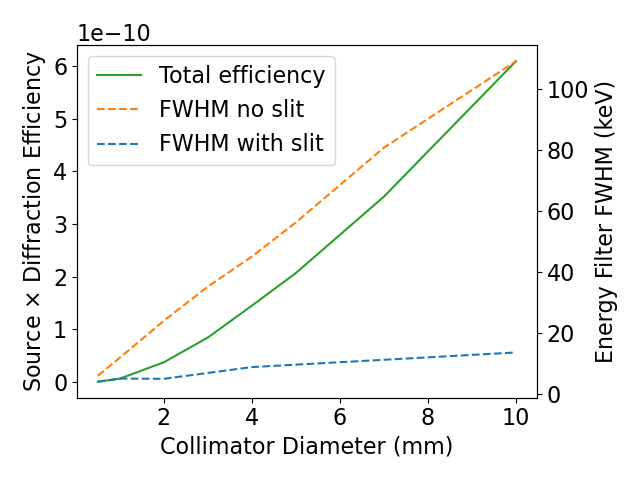}
  \end{subfigure}
  \caption{Impact of varying the collimator aperture for a fixed collimator length of 50 cm, an infinitely wide crystal, and a 300 cm optical element distance. The tradeoff between diffraction efficiency and the collimator efficiency is shown (left). The product of the two efficiency factors and the energy band width of gammas diffracting in the crystal and reaching the detector slit is also shown (right).}
  \label{fig:collimation_study}
\end{figure*}

The energy resolution of the spectrometer can be estimated for on-Rowland geometry using Equation~\ref{eqn:spectralresolution}, where $\omega_D$ is the Darwin width of the crystal~\cite{shadow_crystal_tutorial}.
\begin{equation}
  \frac{\Delta\lambda}{\lambda_0}=\frac{\Delta E}{E_0}=(\Delta_{src}+\omega_D) \cot\theta_{bragg}
  \label{eqn:spectralresolution}
\end{equation}
The higher energy of gammas compared to X-rays means that the Bragg angle will be very low, and significant collimation will be needed to achieve the desired energy resolution. While collimation improves the diffraction efficiency and energy resolution, it risks reducing the overall efficiency by blocking source radiation. Furthermore, Equation~\ref{eqn:spectralresolution} does not account for crystal mosaicity or the effect of detector slit. 

These relationships were explored using SHADOW3 and are demonstrated for the 661 keV gamma in Figure~\ref{fig:collimation_study}. The collimator aperture was varied between 0.25 and 10 mm. The FWHM with and without the slit was determined, as well as the diffracted intensity of the 661 keV gamma. This was done for an infinite plane crystal with a depth of 3 mm. A fixed collimator length of 50 cm was selected to limit the design space while providing ample shielding from undiffracted radiation.

Serpent was used to determine the source efficiency of the pebble and collimator. Gammas were emitted from the fuel kernels and tallied when they passed through the exit aperture of the collimator at full energy. This was simulated for each spectrometer energy. As a variance reduction method, the source direction was constrained to a cone created with a semi-aperture based on the collimator length and radius. A factor of 2 was applied to this angle to capture other valid pathways without excessively reducing computational efficiency. The gamma current was weighted by a correction factor calculated in Equation~\ref{eqn:cone_emission_fraction}, to obtain the source efficiency, $\varepsilon _{source}$. 
\begin{equation}
    \varepsilon_{cone}=\frac{2\pi(1-\cos(\phi))}{4\pi}=\frac{1-\cos(\phi)}{2}
  \label{eqn:cone_emission_fraction}
\end{equation}
This factor is energy dependent because it also captures the effects of self-shielding, and it leads to the largest reduction in signal. The source efficiency and the diffraction efficiency as a function of the collimator aperture are shown in Figure~\ref{fig:collimation_study}.

As the divergence of the source beam decreases, the diffraction efficiency increases, but the source efficiency drops at a faster rate. Meanwhile, the range of energies that diffract in the crystal increases with divergence. However, the width of the energy band passing through the slit increases much more slowly. Thus, it is preferable to use larger apertures to maximize the overall efficiency, while the slit enables adequate energy filtration. An aperture of 5 mm was used for all the final designs.

\begin{figure}[htb]
    \centering
    \includegraphics[width=0.9\columnwidth]{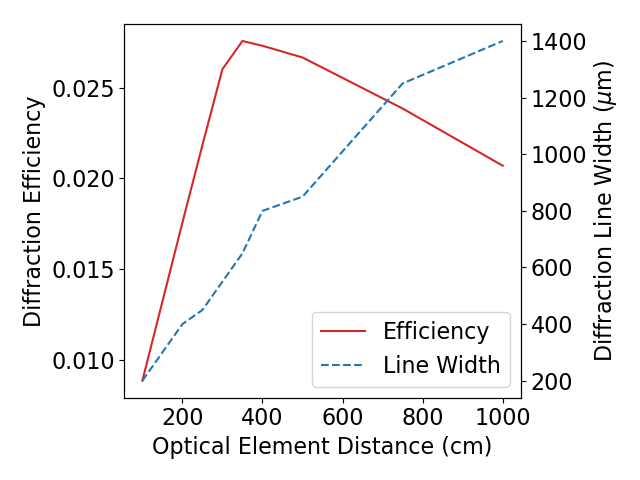}
    \caption{Product of diffraction efficiency and source-collimator geometric efficiency.}
    \label{fig:distance_study_product}
\end{figure}

The impact of the distance between optical elements (i.e. $D_{source}$ and $D_{detector}$) was assessed. For a fixed collimator with an aperture of 2 mm and a length of 50 cm, the distance between elements was varied between 100 cm and 1000 cm. Once again, the planar dimensions of the crystal were infinite with a fixed thickness of 3 mm. Figure~\ref{fig:distance_study_product} shows how distance affects the intensity of the diffracted beam as well as the width of the diffraction line incident on the detector.

To a certain point, increasing the distance improved the diffraction efficiency. This is because the angle of the gammas incident to the crystal lattice becomes more uniform with increasing distance. The inverse square law has a smaller impact on the intensity since the beam is already collimated. The increased angular dispersion also increases the width of the diffraction pattern. Larger distances between optical elements also require an increasingly large crystal in order to capture the entire beam. Both of these parameters affect the footprint and cost of the measurement system. In the absence of space constraints, the best performing element distance of 350 cm was used for all designs.

\begin{figure}[htb]
    \centering
    \includegraphics[width=0.65\columnwidth]{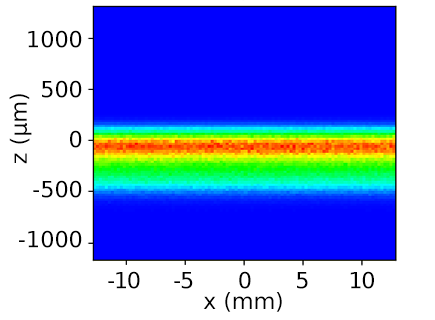}
    \caption{Smearing effect seen on the diffraction pattern for large crystals.}
    \label{fig:distance_study_smearing}
\end{figure}

It was observed that the diffraction pattern became more smeared and less symmetric as the crystal became larger or the source became more divergent. This makes it harder to optimize the slit while maintaining the same throughput. This effect is shown in Figure~\ref{fig:distance_study_smearing}. Asymmetry in the energy filter is not a problem, especially when it is skewed away from interfering gammas and sufficient throughput is achieved for the targeted gamma.

The design process for the other spectrometers was shaped by these observations. The same aperture and optical element distance was used for each setup. In order to increase the throughput, several optimization steps were performed for each target gamma:
\begin{itemize}
\item The mosaicity of the crystal was increased up to the point where the energy filter started to accept major interfering gammas.
\item The crystal size was increased in increments of 10 cm until diminishing returns became apparent.
\item If throughput was still low, the detector size was increased to capture more of the incident diffraction on the detector while improving intrinsic efficiency.
\end{itemize}

\section{Design Evaluation}

\subsection{Final Spectrometer Designs}

The final BCD spectrometer design parameters are listed in Table~\ref{tab:spectrometer_designs}, with a handful of shared parameters listed in Table~\ref{tab:design_constants}. Designs for $^{137m}$Ba feature a 90mm by 90mm HPGe and include two options for crystal mosaicity, 0.02° FWHM and 0.005° FWHM. The 0.02° FWHM design provides a wider filter with more throughput, while the 0.005° FWHM design has a narrower filter with lower efficiency. The wider filter may prove advantageous since the nearby peaks could be resolved with the HPGe detector and included as regression features. All designs for other nuclides use scintillators instead. The $^{239}$Np spectrometer is capable of blocking nearly all interfering gammas while maintaining a very high count rate. The $^{144}$Ce design also achieves near perfect energy filtration, albeit at a much lower count rate than $^{239}$Np. The $^{148m}$Pm setup suffers from both a low count rate and significant interference from adjacent gammas that are too close to be filtered. Targeting a higher energy gamma and using an HPGe detector, as with the $^{137m}$Ba design, may be more effective but expensive.

\begin{table*}[tbp]
\caption{Design parameters and performance of BCD spectrometers tailored for measuring different nuclides.}
\centering
\begin{tabular}{|l|l|l|l|l|l|}
\hline
Nuclide                                                                       & $^{137m}$Ba        & $^{137m}$Ba        & $^{239}$Np         & $^{144}$Ce         & $^{148m}$Pm        \\ \hline
Gamma Energy (keV)                                                            & 661.656       & 661.656       & 106.123       & 133.515       & 414.07        \\ \hline
Detector                                                                      & HPGe 90mm     & HPGe 90mm     & BGO 1x1"      & BGO 2x2"      & BGO 2x2"      \\ \hline
Crystal Plane                                                                 & Si (440) & Si (440) & Si (220) & Si (440) & Si (440) \\ \hline
\begin{tabular}[c]{@{}l@{}}Crystal mosaicity \\ (deg, FWHM)\end{tabular}       & 0.02          & 0.005         & 0.008         & 0.03          & 0.005         \\ \hline
Crystal Dimensions (cm)                                                       & 10x160x0.6    & 10x160x0.6    & 10x120x0.3    & 10x120x0.6    & 10x160x0.6    \\ \hline
\begin{tabular}[c]{@{}l@{}}Crystal Radius of \\ Curvature (cm)\end{tabular}   & 35861.66      & 35861.66      & 11504.31      & 7274.90       & 22442.18      \\ \hline
Bragg Angle (degrees)                                                         & 0.5592        & 0.5592        & 1.7434        & 2.7576        & 0.8936        \\ \hline
Slit Width (mm)                                                               & 0.18          & 0.088         & 0.155         & 0.31          & 0.0775        \\ \hline
Slit Offset (mm)                                                              & -0.02         & -0.026        & -0.0325       & -0.025        & -0.02125      \\ \hline
FWHM with slit (eV)                                                           & 37800         & 18600         & 1680          & 2520          & 8640          \\ \hline
FWHM diffracting (eV)                                                         & 76139         & 77000         & 3900          & 3240          & 28800         \\ \hline
Source Efficiency                                                             & 6.0315E-8    & 6.0315E-8    & 3.6376E-8    & 2.9007E-8    & 5.5206E-8    \\ \hline
Diffraction Efficiency (\%)                                                   & 3.934         & 2.483         & 6.141         & 1.789         & 2.586         \\ \hline
\begin{tabular}[c]{@{}l@{}}Detector Intrinsic \\ Efficiency (\%)\end{tabular} & 38.86         & 38.86         & 96.19         & 97.48         & 78.65         \\ \hline
Average Count Rate (cps)                                                      & 36.88         & 23.28         & 4,298          & 10.12         & 14.85 \\ \hline       
\end{tabular}
\label{tab:spectrometer_designs}
\end{table*}

\begin{table}[htb]
    \caption{Operating assumptions and design constants.}
    \centering
    \begin{tabular}{|l|l|}
        \hline
        Parameter & Value \\ \hline
        Source to crystal distance $D_{source}$ & 350 cm \\ \hline
        Detector to crystal distance $D_{detector}$ & 350 cm \\ \hline
        Measurement time, $t$ & 20 s \\ \hline
        Fuel decay time before measurement & 1.5 d \\ \hline
        Collimator length & 50 cm \\ \hline
        Collimator aperture & 5 mm \\ \hline
    \end{tabular}
    \label{tab:design_constants}
\end{table}

\subsection{Simulating Counts}

For each BCD spectrometer design, the intensity of the diffracted beam incident on the detector surface, $I_{incident}$, was calculated using Equation~\ref{eqn:detector_incident_rate}. 
\begin{equation}
  I_{incident}({\gamma})=S_{\gamma}\cdot\varepsilon_{diffract}(f,d,E)\cdot\varepsilon_{source}(f,d,E)
  \label{eqn:detector_incident_rate}
\end{equation}
The source emission rate for each gamma is $S_{\gamma}$. The source efficiency calculated with Serpent that accounts for pebble self shielding and collimator geometry is ${\varepsilon}_{source}$. The diffraction efficiency calculated using SHADOW3 for a specific gamma is $\varepsilon_{diffract}$.

The intrinsic efficiency of the detector was calculated by using Serpent to model a cylinder of detector material at room temperature with void boundaries~\cite{transport-material-compendium-detwiler}. Source gammas were emitted from a rectangle that matches the slit dimensions incident on the detector's outer surface and were directed inwards. An energy deposition tally in the detector cell was used with a narrowly defined energy bin centered on the gamma energy to only capture photoelectric absorption.

\begin{figure*}[tbp]
  \centering
  \begin{subfigure}[b]{0.41\textwidth}
    \centering
    \includegraphics[width=\textwidth]{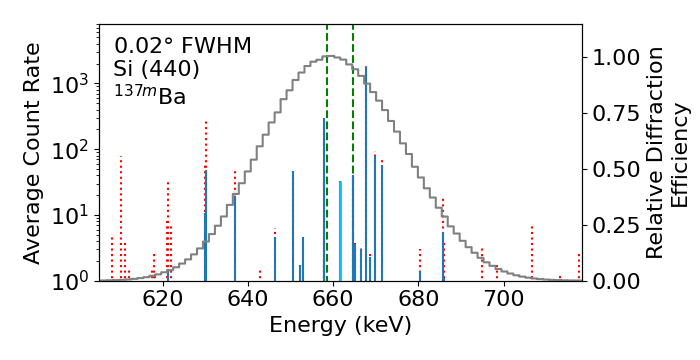}
  \end{subfigure}
  \begin{subfigure}[b]{0.41\textwidth}
    \centering
    \includegraphics[width=\textwidth]{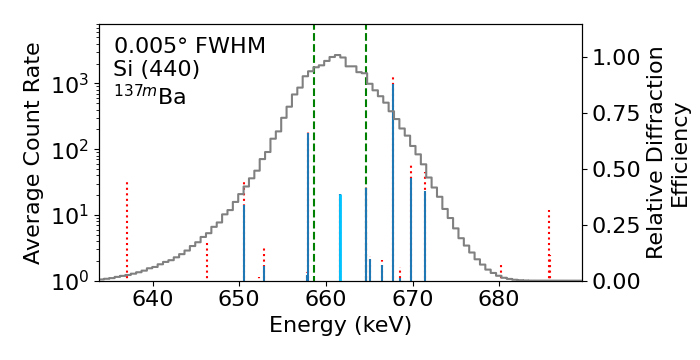}
  \end{subfigure}

  \begin{subfigure}[b]{0.41\textwidth}
    \centering
    \includegraphics[width=\textwidth]{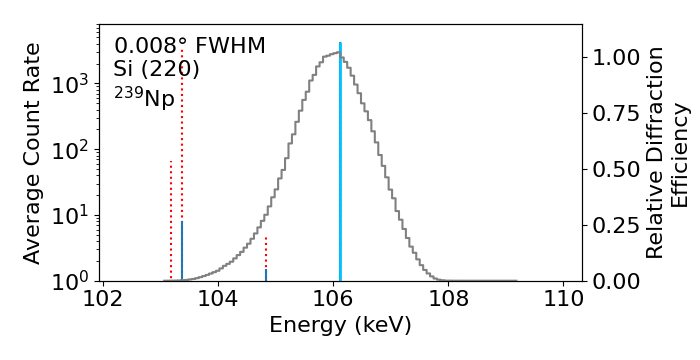}
  \end{subfigure}
  \begin{subfigure}[b]{0.41\textwidth}
    \centering
    \includegraphics[width=\textwidth]{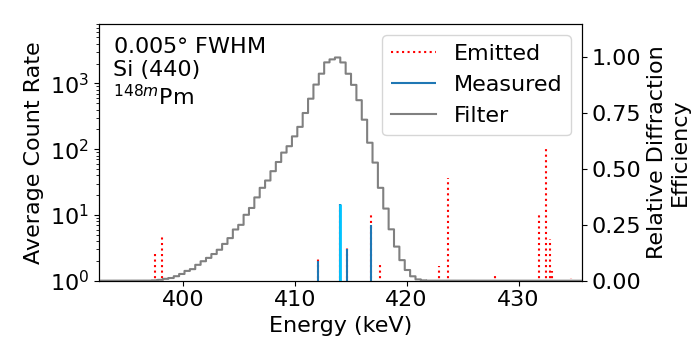}
  \end{subfigure}

  \begin{subfigure}[b]{0.41\textwidth}
    \centering
    \includegraphics[width=\textwidth]{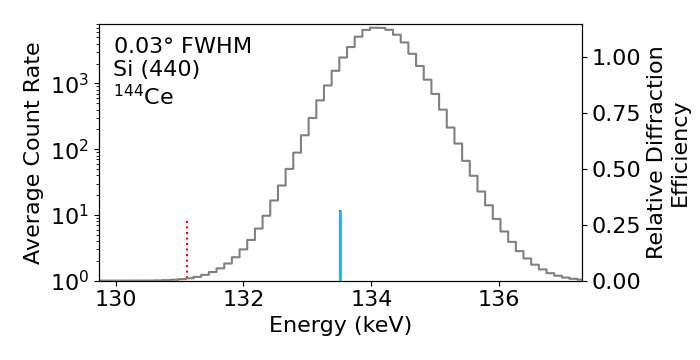}
  \end{subfigure}
  \caption{Example emitted gamma spectra from a pebble after being weighted by source, diffraction, and detector efficiency. The energy filter provided by the BCD spectrometer as calculated in SHADOW3 is overlaid. The red dotted lines represent the intensity of the peak before the relative diffraction efficiency is accounted for. For the $^{137m}$Ba spectrometers, the 1.5$\times$FWHM lines are included around 661 keV. Examples are shown for $^{137m}$Ba (with the coarse filter), $^{137m}$Ba (with the fine filter), $^{239}$Np, $^{138m}$Pm, and $^{144}$Ce.}
  \label{fig:weighted_spectra}
\end{figure*}

Finally, the average count rate in a detector, $C$, was determined. For a detector operated in current mode, this can be calculated using Equation~\ref{eqn:detector_count_rate} where $\Delta E$ includes the entire range of gamma energies that can diffract through the slit. 
\begin{equation}
  C=\varepsilon_{detector}\sum_{\gamma_E}^{\Delta E} I_{incident}({\gamma_E})
  \label{eqn:detector_count_rate}
\end{equation}
With a HPGe operated in pulse mode, the count rates for each peak just become the product of $\varepsilon _{detector}$ and $I_{incident}(\gamma_E)$ for each gamma. It must be verified that close photopeaks are actually resolvable. For an HPGe, a typical FWHM of 2 keV, or 0.3\%, at 661 keV can be assumed. It can also be assumed that peaks separated in energy by 1.5 times the FWHM can be resolved using deconvolution techniques. Thus, gammas that are about 3 keV apart were treated as separate features. Photopeaks that completely masked their neighbors were used over the weaker ones. Ultimately, this meant that the following nearby peaks could be usable as additional features: 667.7 keV from $^{132}$I, 657.9 keV from $^{97}$Nb, 664.6 keV from $^{143}$Ce, 637.0 from $^{131}$I.

\begin{figure*}[htb]
  \centering
  \begin{subfigure}[b]{0.41\textwidth}
    \centering
    \includegraphics[width=\textwidth]{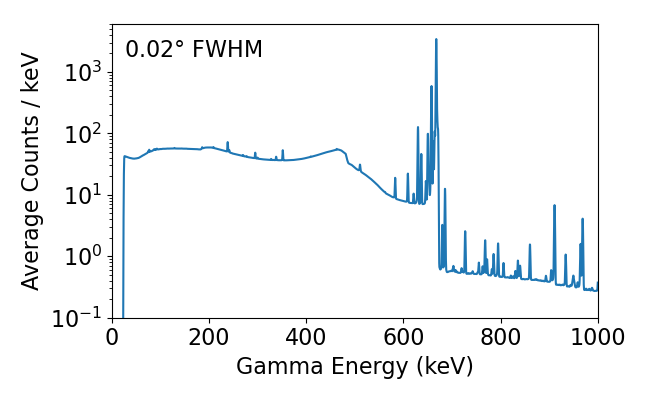}
  \end{subfigure}
  \begin{subfigure}[b]{0.41\textwidth}
    \centering
    \includegraphics[width=\textwidth]{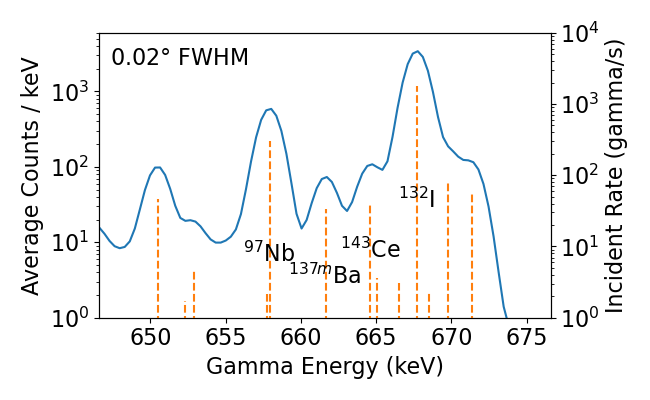}
  \end{subfigure}

  \begin{subfigure}[b]{0.41\textwidth}
    \centering
    \includegraphics[width=\textwidth]{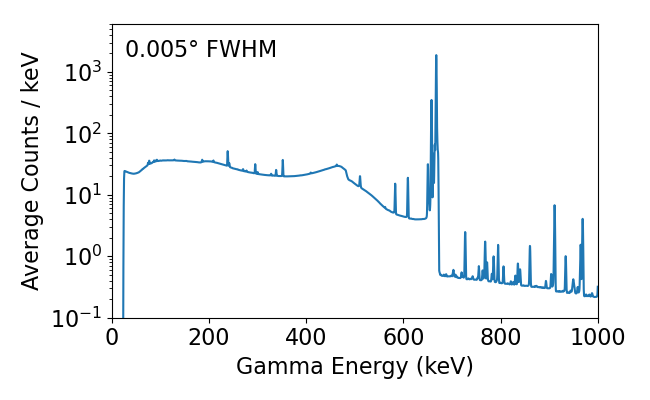}
  \end{subfigure}
  \begin{subfigure}[b]{0.41\textwidth}
    \centering
    \includegraphics[width=\textwidth]{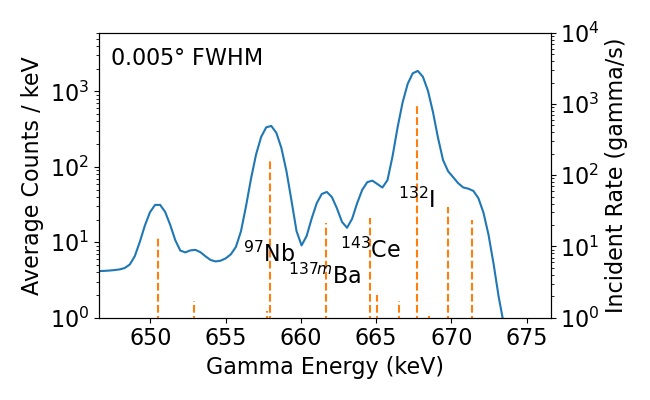}
  \end{subfigure}
  \caption{Simulation of average detected spectrum in GADRAS for pebble using a HPGe detector using a $^{137m}$Ba spectrometer with a crystal mosaicity of 0.02° versus a mosaicity of 0.005°.}
  \label{fig:gadras_example}
\end{figure*}

The average count rate for each spectrometer was calculated from the emission spectrum of each fuel pebble. The spectra of one pebble for each spectrometer is shown in Figure~\ref{fig:weighted_spectra}. The total number of counts over a 20 second interval was sampled from a Poisson distribution with $\lambda=20C$. Thus, synthetic count measurements for each spectrometer and each pebble were generated. 

GADRAS was used to illustrate the observable spectrum in the $^{137m}$Ba detector as well as verify assumptions about peak resolution. GADRAS uses a combination of first-principles calculations and empirical data to generate a realistic measured gamma spectrum~\cite{GADRAS}. A comparison of the average detector response for the two $^{137m}$Ba spectrometer designs is shown in Figure~\ref{fig:gadras_example}. It can be seen that the gammas around 661 keV can be clearly resolved, justifying their use as separate regression features. GADRAS was only used for verification and not for dataset generation. There is no expectation of separating any peaks with scintillators, so the corresponding spectrometers were not simulated in GADRAS.

\subsection{Regression Performance}
\label{subsec:ModelPerformance}

To test the predictive potential of these measurements, a set of RFR models were trained with pebble data using the synthetic measurements as input and the history parameters from HxF as output. Models trained in this way could be used in practice. The coefficient of determination ($R^2$) and mean absolute percent error (MAPE) is shown for these models in Table~\ref{tab:cumulativeresults} and Table~\ref{tab:lastpassresults} for total and last-pass history parameters and nuclide concentrations respectively. The values in the Actual Concentration column refer to the $R^2$ coefficients for the models using the HxF concentrations of $^{137m}$Ba, $^{239}$Pu, $^{144}$Ce, $^{148m}$Pm, and $^{140}$La as input. This represents ideal measurement conditions. The other pairs of columns correspond to models that use the synthetic spectrometer measurements as input. The key difference is whether the coarse filter or fine filter setup is used for the HPGe detector measuring $^{137m}$Ba. Bismuth Germanate (BGO) detectors are used in current mode for all other nuclides due to their relatively low cost and high intrinsic efficiency. The mean average percent error is also included for the synthetic measurement models.

\begin{table*}[tbp]
    \caption{$R^2$ accuracy and MAPE for cumulative history parameters, which are tracked over all of the passes the pebble has taken up to that point.}
    \label{tab:cumulativeresults}
    \centering
\begin{tabular}{|l|l|l|l|l|l|}
\hline
                   & \begin{tabular}[c]{@{}l@{}}$^{137m}$Ba \\ (0.02°) $R^2$\end{tabular} & \begin{tabular}[c]{@{}l@{}}$^{137m}$Ba \\ (0.005°) $R^2$\end{tabular} & \begin{tabular}[c]{@{}l@{}}Actual \\ Concentrations \\ $R^2$\end{tabular} & \begin{tabular}[c]{@{}l@{}}$^{137m}$Ba \\ (0.02°) MAPE\end{tabular} & \begin{tabular}[c]{@{}l@{}}$^{137m}$Ba \\ (0.005°) MAPE\end{tabular} \\ \hline
Burnup             & 0.9953                                                               & 0.9930                                                                & 1.0000                                                                    & 2.28                                                                & 2.70                                                                 \\ \hline
Number of Passes   & 0.9880                                                               & 0.9851                                                                & 0.9994                                                                    & 2.27                                                                & 2.86                                                                 \\ \hline
Residence Time     & 0.9882                                                               & 0.9853                                                                & 0.9994                                                                    & 2.57                                                                & 3.09                                                                 \\ \hline
$^{235}$U Content  & 0.9958                                                               & 0.9947                                                                & 0.9981                                                                    & 3.49                                                                & 3.85                                                                 \\ \hline
$^{239}$Pu Content & 0.8862                                                               & 0.8829                                                                & 0.8883                                                                    & 5.11                                                                & 5.18                                                                 \\ \hline
Thermal Fluence    & 0.9873                                                               & 0.9845                                                                & 0.9924                                                                    & 3.85                                                                & 4.19                                                                 \\ \hline
Fast Fluence       & 0.9811                                                               & 0.9771                                                                & 0.9874                                                                    & 5.32                                                                & 5.79                                                                 \\ \hline
\end{tabular}
\end{table*}

It is clear that using a higher mosaicity crystal for the HPGe spectrometer offers marginally better results due to the improved counting statistics of the $^{137m}$Ba peak. There is very little room for improvement between using synthetic measurements and the "true" HxF concentrations as input. In some cases, the measured spectra actually allow for more accurate prediction due to the extra information available from neighboring peaks around 661 keV in the HPGe setup. This is because only the concentrations of target nuclides are used for the ideal case.

\begin{table*}[tbp]
\caption{$R^2$ accuracy and MAPE for last pass history parameters, which are tracked over the pebble's most recent pass through the core.}
\centering

\begin{tabular}{|l|l|l|l|l|l|}
\hline
                    & \begin{tabular}[c]{@{}l@{}}$^{137m}$Ba \\ (0.02°) $R^2$\end{tabular} & \begin{tabular}[c]{@{}l@{}}$^{137m}$Ba \\ (0.005°) $R^2$\end{tabular} & \begin{tabular}[c]{@{}l@{}}Actual \\ Concentrations \\ $R^2$\end{tabular} & \begin{tabular}[c]{@{}l@{}}$^{137m}$Ba \\ (0.02°) MAPE\end{tabular} & \begin{tabular}[c]{@{}l@{}}$^{137m}$Ba \\ (0.005°) MAPE\end{tabular} \\ \hline
Burnup              & 0.9944                                                               & 0.9943                                                                & 0.9948                                                                    & 1.59                                                                & 1.61                                                                 \\ \hline
Average Radial Path & 0.8793                                                               & 0.8748                                                                & 0.8814                                                                    & 10.84                                                               & 10.94                                                                \\ \hline
Thermal Fluence     & 0.7425                                                               & 0.7338                                                                & 0.8164                                                                    & 3.51                                                                & 3.55                                                                 \\ \hline
Fast Fluence        & 0.9191                                                               & 0.9175                                                                & 0.9270                                                                    & 4.03                                                                & 4.05                                                                 \\ \hline
\end{tabular}
\label{tab:lastpassresults}
\end{table*}

\begin{figure*}[tbp]
  \centering
  \begin{subfigure}[b]{0.38\textwidth}
  \centering
  \includegraphics[width=\textwidth]{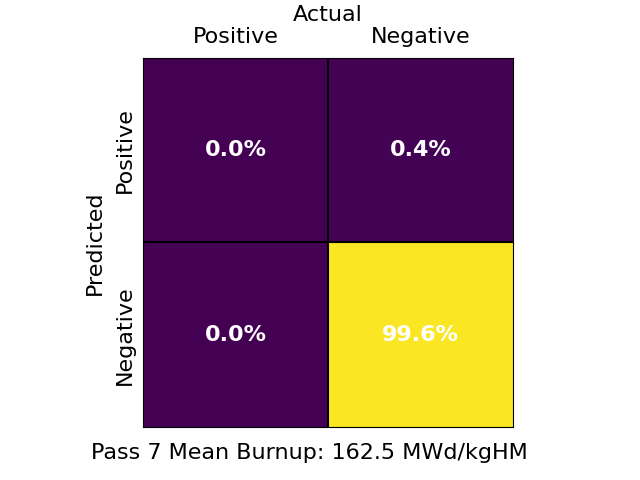}
  \end{subfigure}
  \begin{subfigure}[b]{0.38\textwidth}
    \centering
  \includegraphics[width=\textwidth]{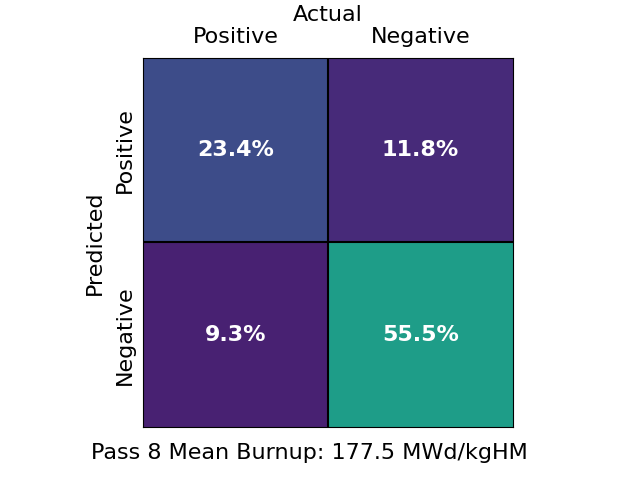}
  \end{subfigure}

  \begin{subfigure}[b]{0.38\textwidth}
  \centering
  \includegraphics[width=\textwidth]{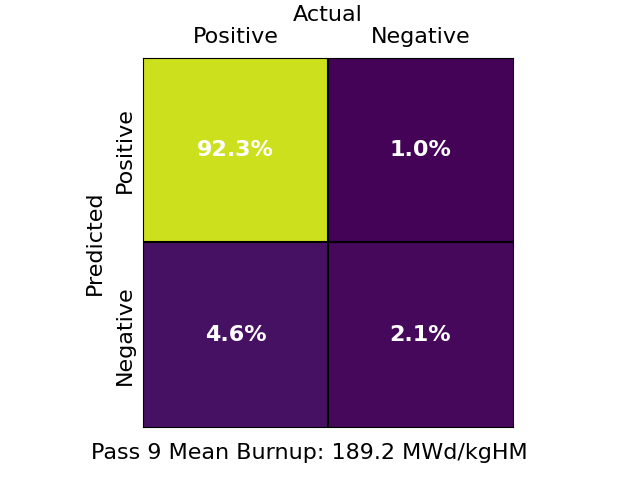}
  \end{subfigure}
  \begin{subfigure}[b]{0.38\textwidth}
    \centering
  \includegraphics[width=\textwidth]{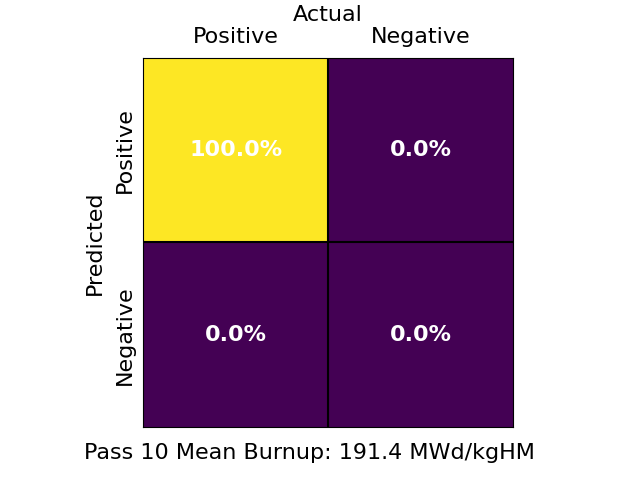}
  \end{subfigure}
  \caption{Confusion matrices showing pebble handling performance based on predicted burnup where crystal mosaicity for the $^{137m}$Ba measuring spectrometer is 0.02° FWHM. Matrix is shown for pass 7, pass 8, pass 9, and pass 10 pebbles.}
  \label{fig:confusion_matrices}
\end{figure*}
 
To illustrate the accuracy of the modeled burnup values in the context of pebble handling performance, confusion matrices were constructed based on the correct classification of pebble burnup relative to the operating discard threshold of 180 MWd/kgHM. That is, pebbles with burnup greater than the threshold are labeled as "positive," while those under the threshold are "negative". These scores were calculated by pass, and are shown in Figure~\ref{fig:confusion_matrices} for passes 7-10. Unsurprisingly, pass 7 and pass 10 pebbles have near perfect accuracy, since almost all of the pebbles in that pass are far away the threshold. The average burnup of pass 8 pebbles is 177.5 MWd/kgHM, which is very close to the threshold of 180 MWd/kgHM. Optimization of the setup should minimize the false positive and false negative values as well as keep them close together. Having a significant skew towards either type of error could eventually have an impact on the reactivity of the core, while close values could compensate for each other.

\section{Gamma Heating Bounding Calculation}
It is crucial for the spectrometer crystal to be kept thermally stable, as large temperature fluctuations can change the diffraction properties of the crystals. Thermal expansion can change the lattice spacing $d$ of the crystal, which can shift the Bragg angle and cause the system to become misaligned. In extreme cases where the crystal is heated unevenly, the crystal can be warped or subject to local strain. Changes in the ambient temperature are most likely to drive these effects. However, gamma heating could potentially play a role as well, and its impact must be quantified.

\begin{figure}[htb]
  \centering
    \includegraphics[width=0.9\columnwidth]{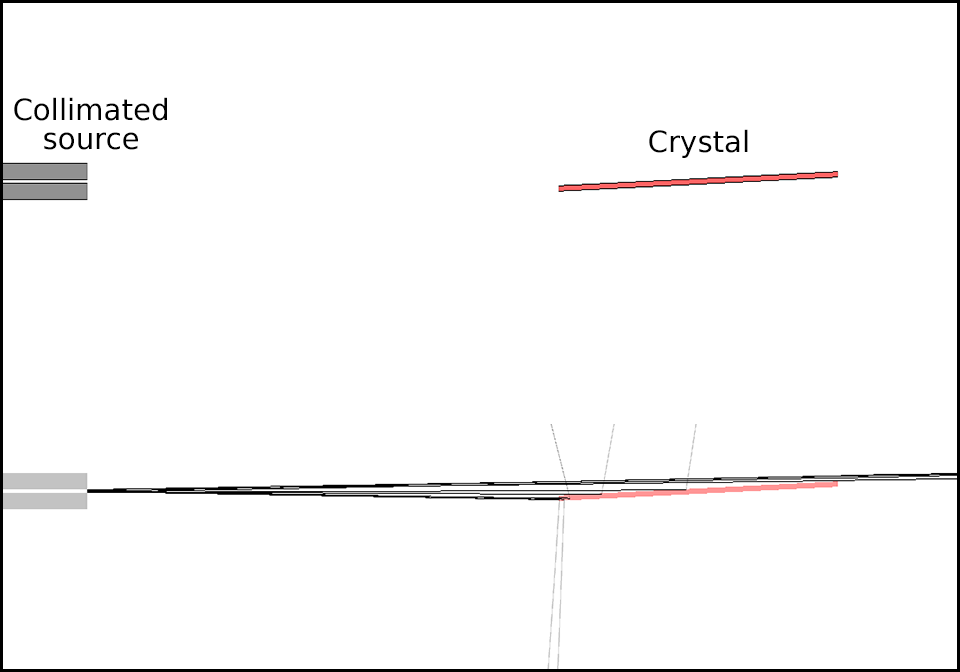}
    \caption{Serpent plot of collimated beam hitting crystal with and without tracked particles, illustrating the spread of the gammas.}
    \label{fig:gamma_heating_model}
\end{figure}

\begin{figure}[htb]
    \centering
    \includegraphics[width=0.9\columnwidth]{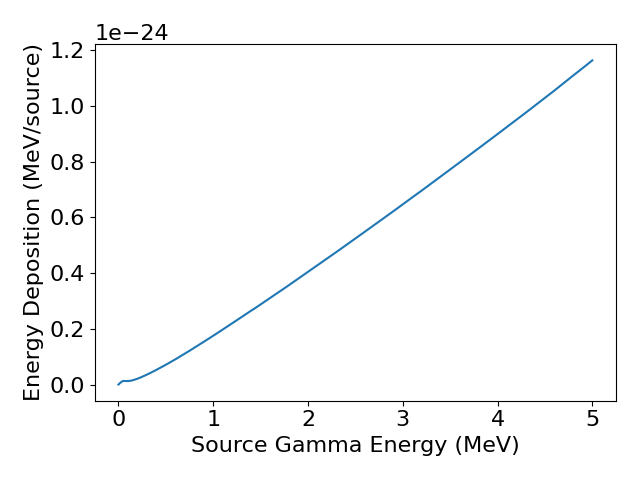}
    \caption{Energy deposition in the crystal as a function of source gamma energy.}
    \label{fig:gamma_heating_curve}
\end{figure}

For the $^{137m}$Ba design, Serpent was used to model the collimated beam hitting the bent crystal, which is approximated as a cuboid. 200 monoenergetic simulations were performed with source energies ranging from 1.125 keV to 5 MeV. The energy deposition per source gamma was tallied across the whole crystal volume. A plot of the gamma heating simulation is shown in Figure~\ref{fig:gamma_heating_model}, and the energy deposition curve is shown in Figure~\ref{fig:gamma_heating_curve}.

Next, the energy deposition from the whole spectrum was calculated. For each energy of gamma emitted from the pebble, the corresponding escape efficiency, $\varepsilon_{escape}$ was interpolated from the curve in Figure~\ref{fig:transmission_curve}, while the energy deposition, $H$, was interpolated from Figure~\ref{fig:gamma_heating_curve}. Equation~\ref{eqn:gamma_heating} was used to calculate the wattage in the crystal due to each gamma. 
\begin{equation}
P=S_\gamma\cdot\varepsilon_{escape}(E_\gamma) \cdot H(E_\gamma) \cdot\varepsilon_{collimator}(f,d)
\label{eqn:gamma_heating}
\end{equation}
The collimator efficiency without any dependence on energy, $\varepsilon_{collimator}$, can be calculated using Equation~\ref{eqn:collimator_efficiency} and using the results at the same energy from Figure~\ref{fig:transmission_curve} and Table~\ref{tab:spectrometer_designs}. Finally, the total gamma heating was determined by summing the wattage for each gamma.
\begin{equation}
\varepsilon_{collimator}(f,d)=\frac{\varepsilon_{source}(f,d,E)}{\varepsilon_{escape}(E)}
\label{eqn:collimator_efficiency}
\end{equation}

For a sample of pebbles analyzed in this way, the heating on the crystal from the full spectrum ranged from $1.4\times10^{-18}$ to $3.0\times10^{-18}$ W. Even if the beam divergence was removed to perfectly focus the gammas onto the middle of the crystal, the maximum heating did not exceed $4.5\times10^{-18}$ W. Thus, gamma heating can be assumed to be negligible compared to ambient effects. 

\section{Conclusion}
BCD spectrometers enable the direct measurement of select fission products in highly radioactive PBR fuel elements by serving as narrow energy filters for gamma rays. Regression analyses using synthetic measurements show strong potential for accurately and rapidly predicting burnup and other fuel pebble properties. Incorporating spectrometers into a BUMS eliminates the need for additional out-of-core cooling, anti-Compton shielding, and extended measurement times. For certain nuclides, this approach also permits the use of lower-cost detectors. Beyond burnup monitoring, BCD spectrometers offer a powerful tool for safeguards applications due to their ability to quantify plutonium production and potentially measure other actinides.

For measuring $^{137m}$Ba/$^{137}$Cs, combining a BCD spectrometer with an HPGe detector was found to be the most effective approach. The spectrometer removes higher-energy Compton scattering and reduces the overall count rate to manageable levels, while the HPGe detector provides sufficient resolution to distinguish the 661 keV peak from nearby peaks. Additional work is needed to evaluate the influence of neighboring gamma emissions on the $^{137m}$Ba peak, particularly Compton scattering from the 667.7 keV line of $^{132}$I. Using GADRAS to generate a more realistic detector response for each pebble and performing peak fitting is one way of doing this.

For measuring $^{239}$Np and $^{144}$Ce, the filter bandwidth is narrow enough to effectively isolate their characteristic gamma lines, making high-density, low-cost detectors such as BGO scintillators practical. There is also potential to employ silicon crystals with the (111) plane, which are inexpensive and readily available due to their widespread use in the semiconductor industry. Measuring $^{148m}$Pm, however, may require a higher-mosaicity crystal paired with an HPGe detector to resolve its dominant 550 keV peak. Because this configuration is more costly, the decision to implement it depends on whether direct plutonium quantification is prioritized over indirect production monitoring. Alternatively, the direct measurement of plutonium X-rays could be explored, although their lower energy and intensity could make doing this rapidly difficult.

Reactor developers interested in using BCD spectrometers would benefit from performing a systemic design parameter study constrained by footprint and crystal costs. Large crystal dimensions were used to maximize performance capabilities. However, this requires stacking many crystals in practice, which could be expensive depending on the manufacturing requirements. Smaller crystal configurations remain feasible, though with reduced count rates. Future work should also investigate non-equilibrium operating conditions, requiring regression models to incorporate additional parameters such as recent reactor power history or pebble circulation rate. This is especially important for properties that correlate with short-term fission history. Finally, detailed shielding simulations are needed to verify that the collimator and detector slit effectively suppress unwanted gamma rays, particularly for detectors operated in current mode.

\section*{Acknowledgments}

This research uses the Savio computational cluster resource provided by the Berkeley Research Computing program at the University of California, Berkeley (supported by the UC Berkeley Chancellor, Vice Chancellor for Research and Chief Information Officer). It was funded by the Nuclear Regulatory Commission.

\bibliography{_bibliography}

@misc{GADRAS,
  author       = {Mitchell, Dean and \& Mattingly, John},
  title        = {Gamma Detector Response and Analysis Software (GADRAS) v. 16.0},
  url          = {https://www.osti.gov/biblio/1231259},
  place        = {United States},
  year         = {2009},
  month        = {12}}

@article{xop,
author = "Sanchez del Rio, Manuel and Perez-Bocanegra, Nicolas and Shi, Xianbo and Honkim{\"{a}}ki, Veijo and Zhang, Lin",
title = "{Simulation of X-ray diffraction profiles for bent anisotropic crystals}",
journal = "Journal of Applied Crystallography",
year = "2015",
volume = "48",
number = "2",
pages = "477--491",
month = "Apr",
doi = {10.1107/S1600576715002782},
url = {https://doi.org/10.1107/S1600576715002782},
}

@inproceedings{shadow3,
author = {F. Cerrina},
title = {{Ray Tracing Of Recent VUV Monochromator Designs}},
volume = {0503},
booktitle = {Application, Theory, and Fabrication of Periodic Structures, DiffractionGratings, and Moire Phenomena II},
editor = {Jeremy M. Lerner},
organization = {International Society for Optics and Photonics},
publisher = {SPIE},
pages = {68 -- 77},
year = {1984},
doi = {10.1117/12.944815},
URL = {https://doi.org/10.1117/12.944815}
}

@inproceedings{shadow_crystal_tutorial,
author = {Manuel Sanchez del Rio},
title = {{Ray tracing simulations for crystal optics}},
volume = {3448},
booktitle = {Crystal and Multilayer Optics},
editor = {Albert T. Macrander and Andreas K. Freund and Tetsuya Ishikawa and Dennis M. Mills},
organization = {International Society for Optics and Photonics},
publisher = {SPIE},
pages = {230 -- 245},
year = {1998},
doi = {10.1117/12.332511},
URL = {https://doi.org/10.1117/12.332511}
}

@article{highcounthpge,
title = {A prototype High Purity Germanium detector for high resolution gamma-ray spectroscopy at high count rates},
journal = {Nuclear Instruments and Methods in Physics Research Section A: Accelerators, Spectrometers, Detectors and Associated Equipment},
volume = {795},
pages = {167-173},
year = {2015},
issn = {0168-9002},
doi = {https://doi.org/10.1016/j.nima.2015.05.053},
url = {https://www.sciencedirect.com/science/article/pii/S0168900215007123},
author = {R.J. Cooper and M. Amman and P.N. Luke and K. Vetter}
}

@article{best-candidate-isotopes-burnup,
title = {Review and characterization of best candidate isotopes for burnup analysis and monitoring of irradiated fuel},
journal = {Annals of Nuclear Energy},
volume = {69},
pages = {278-291},
year = {2014},
issn = {0306-4549},
doi = {https://doi.org/10.1016/j.anucene.2013.12.014},
url = {https://www.sciencedirect.com/science/article/pii/S0306454913006658},
author = {T. Akyurek and L.P. Tucker and S. Usman}
}

@article{synchrotron-two-bcd-monochromators-yamaoka,
author = "Yamaoka, H. and Hiraoka, N. and Ito, M. and Mizumaki, M. and Sakurai, Y. and Kakutani, Y. and Koizumi, A. and Sakai, N. and Higashi, Y.",
title = "{Performance of bent-crystal monochromators for high-energy synchrotron radiation}",
journal = "Journal of Synchrotron Radiation",
year = "2000",
volume = "7",
number = "2",
pages = "69--77",
month = "Mar",
doi = {10.1107/S090904959901691X},
url = {https://doi.org/10.1107/S090904959901691X},
}

@techreport{sandia-pbr-mca-kovacic-report,
  author       = {Kovacic, Donald N. and Gibbs, Philip and Hu, Jianwei and Hartanto, Donny and Ball, Cory and Mcelroy Jr, Robert and Luciano, Nicholas and Hunneke, Rachel and Pham, Tom and Wieselquist, William},
  title        = {Nuclear Material Control \& Accounting for Pebble Bed Reactors (FY 2023 Summary Report)},
  institution  = {Oak Ridge National Laboratory (ORNL), Oak Ridge, TN (United States)},
  doi          = {10.2172/2434394},
  url          = {https://www.osti.gov/biblio/2434394},
  place        = {United States},
  year         = {2023},
  month        = {11}}

@techreport{transport-material-compendium-detwiler,
  author       = {Detwiler, Rebecca S. and McConn, Ronald J. and Grimes, Thomas F. and Upton, Scott A. and Engel, Eric J.},
  title        = {Compendium of Material Composition Data for Radiation Transport Modeling},
  institution  = {Pacific Northwest National Laboratory (PNNL), Richland, WA (United States)},
  doi          = {10.2172/1782721},
  url          = {https://www.osti.gov/biblio/1782721},
  place        = {United States},
  year         = {2021},
  month        = {04}}

@phdthesis{ian-dissertation,
author={Kolaja,Ian T.},
year={2025},
title={Advanced Methods for Measuring and Analyzing Pebble Bed Reactors},
journal={ProQuest Dissertations and Theses},
pages={205},
note={Copyright - Database copyright ProQuest LLC; ProQuest does not claim copyright in the individual underlying works; Last updated - 2025-10-03},
school={University of California, Berkeley},
isbn={9798293893676},
language={English},
url={https://www.proquest.com/dissertations-theses/advanced-methods-measuring-analyzing-pebble-bed/docview/3256605007/se-2},
}

@article{germanium-bent-monochromator-tungsten-levels-dumond,
title = {A germanium bent-crystal monochromator for nuclear spectroscopy},
journal = {Nuclear Instruments and Methods},
volume = {16},
pages = {17-28},
year = {1962},
issn = {0029-554X},
doi = {https://doi.org/10.1016/0029-554X(62)90094-0},
url = {https://www.sciencedirect.com/science/article/pii/0029554X62900940},
author = {E.J. Seppi and H. Henrikson and F. Boehm and J.W.M. Dumond}
}

@article{mosaic-crystal-monochromators-freund,
title = {Mosaic crystal monochromators for synchrotron radiation instrumentation},
journal = {Nuclear Instruments and Methods in Physics Research Section A: Accelerators, Spectrometers, Detectors and Associated Equipment},
volume = {266},
number = {1},
pages = {461-466},
year = {1988},
issn = {0168-9002},
doi = {https://doi.org/10.1016/0168-9002(88)90430-5},
url = {https://www.sciencedirect.com/science/article/pii/0168900288904305},
author = {Andreas K. Freund}
}

@misc{nudat3,
  author       = {National Nuclear Data Center},
  title        = {NuDat 3.0 Database},
  year         = {2025},
  url = {https://www.nndc.bnl.gov/nudat3/},
  note         = {Brookhaven National Laboratory, U.S. Department of Energy},
  urldate = {2025-8-15}
}

@article{nuclear-data-a-239,
title = {Nuclear Data Sheets for A = 239},
journal = {Nuclear Data Sheets},
volume = {122},
pages = {293-376},
year = {2014},
issn = {0090-3752},
doi = {https://doi.org/10.1016/j.nds.2014.11.003},
url = {https://www.sciencedirect.com/science/article/pii/S0090375214006693},
author = {E. Browne and J.K. Tuli}
}

@article{nuclear-data-a-148,
title = {Nuclear Data Sheets for A = 148},
journal = {Nuclear Data Sheets},
volume = {117},
pages = {1-229},
year = {2014},
issn = {0090-3752},
doi = {https://doi.org/10.1016/j.nds.2014.02.001},
url = {https://www.sciencedirect.com/science/article/pii/S009037521400012X},
author = {N. Nica}
}

@article{nuclear-data-a-140,
    author = "Nica, N.",
    title = "{Nuclear Data Sheets for A=140}",
    doi = "10.1016/j.nds.2018.11.002",
    journal = "Nucl. Data Sheets",
    volume = "154",
    pages = "1--403",
    year = "2018"
}

@article{nuclear-data-a-144,
author = {Sonzogni, A.},
year = {2001},
month = {07},
pages = {599-762},
title = {Nuclear Data Sheets for A = 144},
volume = {93},
journal = {Nuclear Data Sheets - NUCL DATA SHEETS},
doi = {10.1006/ndsh.2001.0015}
}

@article{nuclear-data-a-137,
title = {Nuclear Data Sheets for A = 137},
journal = {Nuclear Data Sheets},
volume = {108},
number = {10},
pages = {2173-2318},
year = {2007},
issn = {0090-3752},
doi = {https://doi.org/10.1016/j.nds.2007.09.002},
url = {https://www.sciencedirect.com/science/article/pii/S0090375207000804},
author = {E. Browne and J.K. Tuli}
}

@article{gamma-spectroscopy-first-85-years-deslattes,
author = {Deslattes, Richard},
year = {2000},
month = {02},
pages = {},
title = {High Resolution $\gamma$-Ray Spectroscopy: The First 85 Years},
volume = {105},
journal = {Journal of Research of the National Institute of Standards and Technology},
doi = {10.6028/jres.105.002}
}

@techreport{areva-readiness-study,
  title        = {Pebble Bed Reactor Technology Readiness Study},
  author       = {AREVA NP Inc.},
  institution  = {AREVA NP Inc.},
  type         = {Technical Data Record},
  number       = {12-9151714-000},
  address      = {USA},
  year         = {2010},
  month        = {Oct},
  url          = {https://art.inl.gov/NGNP/NEAC%202010/Pebble%20Bed%20Reactor%20Technology%20Readiness%20Study%20-%20AREVA.pdf},
  note         = {Prepared under BEA Contract No.\ 000 75310; issued 10/18/2010}
}

@misc{tree-models-outperform-deep-learning-tabular-grinsztajn,
      title={Why do tree-based models still outperform deep learning on tabular data?}, 
      author={Léo Grinsztajn and Edouard Oyallon and Gaël Varoquaux},
      year={2022},
      eprint={2207.08815},
      archivePrefix={arXiv},
      primaryClass={cs.LG},
      url={https://arxiv.org/abs/2207.08815}, 
}

@book{xray-diffraction-theory-book-zachariasen,
  title={Theory of X-ray Diffraction in Crystals},
  author={Zachariasen, W.H.},
  isbn={9780486683638},
  lccn={lc94034490},
  series={Dover classics of science and mathematics},
  url={https://books.google.com/books?id=Ja-KuQAACAAJ},
  year={1994},
  publisher={Dover}
}

@article{bcd-solar-flare-nasa-sylwester,
   title={A Unique Resource for Solar Flare Diagnostic Studies: The SMM Bent Crystal Spectrometer},
   volume={894},
   ISSN={1538-4357},
   url={http://dx.doi.org/10.3847/1538-4357/ab86ba},
   DOI={10.3847/1538-4357/ab86ba},
   number={2},
   journal={The Astrophysical Journal},
   publisher={American Astronomical Society},
   author={Sylwester, J. and Sylwester, B. and Phillips, K. J. H. and Kępa, A. and Rapley, C. G.},
   year={2020},
   month=may, pages={137} }

@software{zenodo-code-archive,
  author       = {iankolaja},
  title        = {iankolaja/pbr-bcd-spectrometer-analysis:
                   v1.0-eprint-archive
                  },
  month        = oct,
  year         = 2025,
  publisher    = {Zenodo},
  version      = {v1eprint},
  doi          = {10.5281/zenodo.16873613},
  url          = {https://doi.org/10.5281/zenodo.16873613},
}

@dataset{hxf-count-data,
  author       = {Kolaja, Ian and
                  Jantzen, Ludovic},
  title        = {HxF Depletion Data \& Synthetic Count Rates for
                   "Burnup Measurement using Bent Crystal
                   Spectrometer for Pebble Bed Reactors"
                  },
  month        = oct,
  year         = 2025,
  publisher    = {Zenodo},
  doi          = {10.5281/zenodo.17308467},
  url          = {https://doi.org/10.5281/zenodo.17308467},
}

@inproceedings{bcd-plutonium-spectrometer-goodsell,
  author    = {A. V. Goodsell and W. S. Charlton},
  title     = {Bent-Crystal Spectrometer Analyzing Plutonium K X-Rays for Applications in Nuclear Forensics},
  booktitle = {Proceedings of the Institute of Nuclear Materials Management Annual Meeting},
  year      = {2011},
  note      = {INMM Annual Meeting},
}

@article{analysis-of-peb-bu-in-PBR,
title = {Analysis of the pebble burnup profile in a pebble-bed nuclear reactor},
journal = {Nuclear Engineering and Design},
volume = {345},
pages = {233-251},
year = {2019},
issn = {0029-5493},
doi = {https://doi.org/10.1016/j.nucengdes.2019.01.030},
url = {https://www.sciencedirect.com/science/article/pii/S0029549318310525},
author = {Yushi Tang and Liguo Zhang and Qiuju Guo and Bing Xia and Zaizhe Yin and Jianzhu Cao and Jiejuan Tong and Chris H. Rycroft}
}

@article{hxf,
	Author = {Y. Robert and T. Siaraferas and M. Fratoni},
	Journal = {Scientific Reports},
	Number = {12711},
	Title = {Hyper-fidelity depletion with discrete motion for pebble bed reactors},
	Volume = {13},
	Year = {2023}}

@article{serpent,
	Author = {J. Leppanen and et al},
	Journal = {Annals of Nuclear Energy},
	Title = {The Serpent Monte Carlo code: Status, development and applications in 2013},
	Volume = {82},
    Pages = {142--150},
	Year = {2015}}

@article{scikit,
author = {Pedregosa, Fabian and Varoquaux, Gael and Gramfort, Alexandre and Michel, Vincent and Thirion, Bertrand and Grisel, Olivier and Blondel, Mathieu and Prettenhofer, Peter and Weiss, Ron and Dubourg, Vincent and Vanderplas, Jake and Passos, Alexandre and Cournapeau, David and Brucher, Matthieu and Perrot, Matthieu and Duchesnay, Edouard and Louppe, Gilles},
year = {2012},
month = {01},
pages = {},
title = {Scikit-learn: Machine Learning in Python},
volume = {12},
journal = {Journal of Machine Learning Research}
}

@article{endf,
title = {ENDF/B-VII.1 Nuclear Data for Science and Technology: Cross Sections, Covariances, Fission Product Yields and Decay Data},
journal = {Nuclear Data Sheets},
volume = {112},
number = {12},
pages = {2887-2996},
year = {2011},
note = {Special Issue on ENDF/B-VII.1 Library},
issn = {0090-3752},
doi = {https://doi.org/10.1016/j.nds.2011.11.002},
url = {https://www.sciencedirect.com/science/article/pii/S009037521100113X},
author = {M.B. Chadwick and M. Herman and P. Obložinský and M.E. Dunn and Y. Danon and A.C. Kahler and D.L. Smith and B. Pritychenko and G. Arbanas and R. Arcilla and R. Brewer and D.A. Brown and R. Capote and A.D. Carlson and Y.S. Cho and H. Derrien and K. Guber and G.M. Hale and S. Hoblit and S. Holloway and T.D. Johnson and T. Kawano and B.C. Kiedrowski and H. Kim and S. Kunieda and N.M. Larson and L. Leal and J.P. Lestone and R.C. Little and E.A. McCutchan and R.E. MacFarlane and M. MacInnes and C.M. Mattoon and R.D. McKnight and S.F. Mughabghab and G.P.A. Nobre and G. Palmiotti and A. Palumbo and M.T. Pigni and V.G. Pronyaev and R.O. Sayer and A.A. Sonzogni and N.C. Summers and P. Talou and I.J. Thompson and A. Trkov and R.L. Vogt and S.C. {van der Marck} and A. Wallner and M.C. White and D. Wiarda and P.G. Young}
}

@article{gfhr,
title = {Neutronics, thermal-hydraulics, and multi-physics benchmark models for a generic pebble-bed fluoride-salt-cooled high temperature reactor (FHR)},
journal = {Nuclear Engineering and Design},
volume = {384},
pages = {111461},
year = {2021},
issn = {0029-5493},
doi = {https://doi.org/10.1016/j.nucengdes.2021.111461},
url = {https://www.sciencedirect.com/science/article/pii/S0029549321004131},
author = {Nader Satvat and Fatih Sarikurt and Kevin Johnson and Ian Kolaja and Massimiliano Fratoni and Brandon Haugh and Edward Blandford}
}

\end{document}